\documentclass[review]{elsarticle}

\usepackage{tikz}
\graphicspath{{images/}}
\usepackage{amsmath}
\usepackage{latexsym, amssymb}
\usepackage{graphicx}
\usepackage{amsmath, amsbsy}
\usepackage{amsopn, amstext}
\usepackage{ifpdf,hyperref}
\usepackage{cancel, color}
\usepackage{epstopdf}
\usepackage[all]{xy}

\def\lplus{\,\rotatebox[]{-90}{$\pm$}\,}

\def\lvminus{\vec{\vdash}}

\def\lvplus{\vec{\lplus}}

\def\J{{\bf 1}}

\def\sign{sign}

\DeclareMathOperator{\rank}{rank}

\DeclareMathOperator{\lcm}{lcm}
\def\cal{\mathcal}

\def\pa{\partial}

\def\ra{\rightarrow}

\def\lra{\leftrightarrow}

\def\w{\wedge}
\def\v{\vee}

\def\a{\alpha}
\def\b{\beta}
\def\d{\delta}

\def\D{\Delta}

\def\0{{\bf 0}}

\def\A{\left<A\right>}

\newcommand{\R}{{\mathbb R}}

\newcommand{\Z}{{\mathbb Z}}

\def\dsum{\mathop{\sum}\limits}

\newtheorem{thm}{Theorem}[section]
\newtheorem{dfn}[thm]{Definition}
\newtheorem{prp}[thm]{Proposition}
\newtheorem{exa}[thm]{Example}
\newtheorem{lem}[thm]{Lemma}
\newtheorem{cor}[thm]{Corollary}
\newtheorem{rem}[thm]{Remark}
\newtheorem{alg}[thm]{Algorithm}

\begin{document}
	
\begin{frontmatter}

\title{From Dimension-Free Manifolds to Dimension-Varying Control Systems\tnoteref{footnoteinfo}}

	\tnotetext[footnoteinfo]{This work is supported partly by the National Natural Science Foundation of China (NSFC) under Grants 62073315 and 61733018. Corresponding author: Zhengping Ji.}
	
\author[AMSS,STPC]{Daizhan Cheng}\ead{dcheng@iss.ac.cn}
\author[AMSS,CAS]{Zhengping Ji}\ead{jizhengping@amss.ac.cn}

\address[AMSS]{Key Laboratory of Systems and Control, Academy of Mathematics and Systems Science, Chinese Academy of Sciences, Beijing 100190, P.R.China}
\address[STPC]{Research Center of Semi-tensor Product of Matrices, Theory and Applications, Liaocheng University, Lianocheng, P.R. China}
\address[CAS]{School of Mathematical Sciences, University of Chinese Academy of Sciences, Beijing 100049, P.R.China}

%


\begin{abstract}
Starting from the vector multipliers, the inner product, norm, distance, as well as addition of two vectors of different dimensions are proposed, which makes the spaces into a topological vector space, called the Euclidean space of different dimension (ESDD). An equivalence is obtained via distance. As a quotient space of ESDDs w.r.t. equivalence, the dimension-free Euclidean spaces (DFESs) and dimension-free manifolds (DFMs) are obtained, which have  bundled vector spaces as its tangent space at each point. Using the natural projection from a ESDD to a DFES, a fiber bundle structure is obtained, which has ESDD as its total space and DFES as its base space.  Classical objects in differential geometry, such as smooth functions, (co-)vector fields, tensor fields, etc., have been extended to the case of DFMs with the help of projections among different dimensional Euclidean spaces. Then the dimension-varying  dynamic systems (DVDSs) and dimension-varying control systems (DVCSs) are presented, which have DFM as their state space. The realization, which is a lifting of DVDSs or DVCSs from DFMs into ESDDs, and the projection of DVDSs or DVCSs from ESDDs onto DFMs are investigated.
\end{abstract}


\end{frontmatter}

\section{Introduction}

Dimension-varying  dynamic systems (DVDSs) and dimension-varying control systems (DVCSs) exist widely in nature and man-made equipments or environments. For instance, on the internet users are joining and withdrawing frequently. In a biological system, cells are producing and dying from time to time. Some man-made mechanical systems are also of varying dimensions. For instance, the docking and undocking of spacecrafts \cite{yan14,jia14}; the connecting and disconnecting of vehicle clutch systems while speed changes \cite{che20a}. The DVDS models are also used for specious population dynamics \cite{xu81,hua95}.

Another interesting phenomenon which stimulates our interest is: a geometrical object, or a complex system,  may be described by models of different dimensions. For instance, in power systems a single generator can be modeled as a $2$, $3$, or $5$, $6$, or even $7$-dimensional dynamic system \cite{mac97}. In contemporary physics, the sting theory assumes the dynamics of strings to be the model for universe of time-space. But this model may have dimension 4 (special relativity), 5 (Kaluza-Klein theory), 10 (type 1 string), 11 (M-theory) or even 26 (Bosonic model) \cite{kak99}. One observes from this phenomenon that two models with different dimensions might be very similar or even equivalent. In other words, dimension-varying model may be proper to describe such dynamics.

So far, a classical way to deal with DVDSs and DVCSs is switching \cite{yan14}. This approach ignores the dynamics of the system during the dimension-varying process. In practice, the transient period may be long enough so that the dynamics during this process is not ignorable.  For instance, automobile clutch takes about $1$ second to complete a connection or separation action; docking and undocking of spacecrafts take even longer. Not to mention that some processes might be continuously dimension-varying. In the latter cases, switching is almost meaningless.

To our best knowledge, there are few proper theories in existing mathematics to model DVDSs and DVCSs. In ordinary or partial differential equations or difference equations, only dynamic models of fixed dimensions can be treated. To provide a proper model for formulating DVDSs and DVCSs, a new framework should be created.

The purpose of this paper is twofold. One is to build the dimension-free Euclidean spaces (DFESs) and dimension-free manifolds (DFMs), which provides a  mathematical framework for DVDSs or DVCSs. The DFM is a completely new ``manifold", where each point has its own dimension. It is the base space of DFES while the total space of DFES is considered as the tangent space of the DFM. It makes the ``state space" of the DVDSs or DVCSs a mathematically well-posed geometrical object; The other issue is to use the geometric structure of DFMs to model, analyze and/or design controls for DVDSs and DVCSs, either linear or nonlinear, are unambiguously defined.

This paper is a follow-up research of our previous works. In \cite{che19,che19b} the dimension-free matrix theory and mix-dimensional vector spaces have been proposed and investigated. As an application of dimension-free matrix theory, the dimension-varying  linear (control) systems have been investigated \cite{che20a}. The basic concept used there was the equivalence of vectors of different dimensions \cite{che20}. This idea is also one of the key techniques in this paper.

To build up a theory of dimension-free manifolds, the key issue is to construct a connected topological space allowing the dimensions of the points in it to vary. We first construct an equivalence relation on the ESDD ${\cal V}:=\bigcup_{n=1}^{\infty}{\cal V}$, (equivalently, $\R^{\infty}:=\bigcup_{n=1}^{\infty}\R^n$) and take the quotient space $\Omega$ as the model space, which is a topological vector space. it is called ``dimension-free" Euclidean space, which means each vector $\bar{x}\in \Omega$ has its dimension $1\leq \dim(\bar{x})< \infty$, but this dimension varies from point to point. We refer to \cite{che19,che19b} for details. Though these results have been discussed before, they are re-organized systematically in Section \ref{s2} here. Then the projection of a vector to a certain Euclidian space is recalled, which was firstly proposed in \cite{che20a} and was discussed in detail by \cite{fen21,fen21b,zha21,zha21b}. Then the cross-dimensional linear control systems are investigated as S-systems (semi-group systems). 
%

For each point $\bar{x}\in\Omega$ a neighborhood coordinate bundle is proposed. Using neighborhood coordinate bundles, differentiable structure is obtained. Unlike the classical differential manifold, this differential structure poses on each point a bundled Euclidean spaces of different dimensions as its ``tangent space". A DFM is a fiber bundle locally homeomorphic to a coordinate neighborhood bundle of a DFES.

Over a DFES (or DFM), the smooth ($C^r$) functions, vector fields, co-vector fields, distributions, co-distributions, tensor fields, etc. are proposed. The integral curves of vector fields, or integral manifolds of distributions, are also properly defined. In a word, a dimension-free differential geometric structure is proposed for DFMs. Dimension-free tensor fields are also introduced, and using degree-2 covariant tensor fields, dimension-free Riemannian manifolds and symplectic manifolds are also proposed.

With these geometric constructions, this paper attempts to explore the dynamics and control of DVDSs. The basic idea is as follows: Projection and lift connect the DVDSs on DFMs (as the base space) with the DVDSs on DFESs (as the total space): lifting the trajectory of a DVDS (or DVCS) to ESDD, a set of trajectories over ESDD are obtained. Using them a dimension-varying trajectory of the DVDS (or DVCS) can be constructed. Conversely, if a classical dynamic (control) system is defined on a Euclidean space, it can also be projected to DFES via natural projection. Therefore, the fiber bundle $(\R^{\infty}, \Pr, \Omega)$ provides the bearing state space for both DVDSs and DVCSs. In addition to general DVDSs, particular attention has been paid to the dynamics of the transient process of classical dimension-varying systems, which have invariant dimensions except during the transient period.


The STP of matrices was proposed by the author and his colleagues \cite{che11,che12}. It was the fundamental method in previous works \cite{che19,che19b,che20a} on DVDSs and DVCSs. It is also a basic tool in this paper, where the default matrix product is assumed to be STP. We refer to \cite{che11,che12} for notions and basic results.

Before ending this section, lists of notations and abbreviations are provided as follows.

\begin{itemize}
\item[(1)] List of Notations:
\end{itemize}

\begin{itemize}

\item $\R$: set of real numbers.
	
\item ${\cal M}_{m\times n}$: set of $m\times n$ dimensional real matrices.

\item $a\v b$: the least common multiple of two positive integers $a$ and $b$. 

\item $a\w b$: the greatest common divisor of two positive integers $a$ and $b$.

%
	
%

\item $\lvplus$: (left) vector addition.

\item $\lvminus$: (left) vector subtraction.

\item $\J_n$: ${\underbrace{[1,\cdots,1]}_n}^T$; $\J_{m\times n}$: $m\times n$ matrices with all entries being $1$.

\item $\d_k^i$: The $i$-th column of identity matrix $I_k$. $\d_k^0$ is for a zero vector of dimension $k$.

\item $\lra$: vector equivalence.

\item $\Omega:=\R^{\infty}/\lra$.

\end{itemize}

\vskip 2mm

\begin{itemize}
\item[(2)] List of  Abbreviations:
\end{itemize}

\begin{itemize}
\item ESDD: Euclidean space of different dimensions.
\item DFES: dimension-free Euclidean space.
\item DFM: dimension-free manifold.
\item DFEB: dimension-free Euclidean bundle.
\item DFRM: dimension-free Riemannian manifold.
\item DVDS: dimension-varying dynamic system.
\item DVCS: dimension-varying control system.
\end{itemize}
\vskip 2mm

The rest of this paper is organized as follows. The ESDDs are investigated in Section \ref{s2}. The inner product, norm, distance, topologies  and equivalence relations on them are introduced. In Section \ref{s3} a vector space structure is given to ESDDs to form the DFESs using equivalence among vectors of different dimensions. Using the natural projection from ESDDs to DFES, a fiber bundle structure is obtained. Section \ref{s4} considers the projections among Euclidian Spaces of different dimensions. First, the projection of a vector onto another vector, which may have different dimensions, is proposed. Using the coordinates from ESDD, smooth functions over DFESs are constructed. Then the least square approximation of a linear (control) system is considered. The DFMs are considered in Section \ref{s5}. After DFESs are endowed with a differential structure, the (co)-vector fields, (co)-distributions, and the integral curses of vector fields over them are proposed and investigated.
In Section \ref{s6}, the tensor fields over DFMs are constructed first. Using proper symmetric and skew-symmetric covariant tensor fields, the dimension-free Riemannian manifold and dimension-free symplectic manifold are constructed respectively. As an application, Section \ref{s7} considers DVDSs and DVCSs. First, the projection of a nonlinear (control) system on an Euclidean space onto another Euclidean space of different dimension is proposed. Then the nonlinear (control) systems over DFESs are considered, which is then used to model dimension-varying nonlinear (control) systems over ESDDs. Finally, the control problems of dimension-varying linear and nonlinear systems are considered in principle. Section \ref{s8} contains some concluding remarks. First, the construction of DFES (DFM) is summarized step by step. Then the modeling and control design of DVDSs are also summarized. Finally, a conjecture is presented, which claims that DFESs (DFMs) might be used as the framework for string theory.

\section{Euclidean Space of Different Dimensions (ESDDs)}\label{s2}

In this section we introduce the notion of ESDDs, which is constructed by choosing $\{\J_n\}_{n=1,2,\cdots}$ as vector multipliers \cite{che20}.

\subsection{Mix-Dimensional Sets and Mix-Dimensional Vector Spaces}

Consider an $n$ dimensional real vector space, denoted by ~${\cal V}_n$. For simplicity, one can take ${\cal V}_n=\R^n$.   To construct mix-dimensional state space, the set of mix-dimensional vectors, called  ESDD, is defined as
$$
{\cal V}:=\bigcup_{n=1}^{\infty}{\cal V}_n.
$$
We may view $\cal V$ as $\R^{\infty}:=\bigcup_{n=1}^{\infty}\R^n$ since they are isomorphic.

First, we define ``addition" and ``scalar product" over ${\cal V}$ to turn it into a pseudo vector space.

\begin{dfn}\label{dpr.1.1}
\begin{itemize}
\item[(i)] Let $x\in {\cal V}_m\subset {\cal V}$, $r\in \R$. Then the scalar product is defined as follows:
\begin{align}\label{pr.1.1}
r\times x:=rx\in {\cal V}_m.
\end{align}

\item[(ii)] Let $x\in {\cal V}_m$, $y\in {\cal V}_n$, and $t=m\v n$ be the least common multiple of $m$ and $n$. Then the addition of $x$ and $y$ is defined as follows:
\begin{align}\label{pr.1.2}
x\lvplus y:=(x\otimes \J_{t/m})+(y\otimes \J_{t/n})\in {\cal V}_t.
\end{align}

Correspondingly, the subtraction of $y$ from $x$ is defined as $x\lvminus y:=x\lvplus (-y)$.
\end{itemize}
\end{dfn}

\begin{prp}\label{ppr.1.4}
Set ${\cal V}$ with scalar multiplication defined as in (\ref{pr.1.1}), addition as in (\ref{pr.1.2}) is a pseudo-vector space \cite{abr78}, where the set of zero elements is
$$
{\bf 0}:=\{[\underbrace{0,0,\cdots,0}_n]^T\;|\;n=1,2,\cdots\}.
$$
\end{prp}

%

\begin{rem}\label{rpr.1.2} For notational ease, when $x\in {\cal V}_n$ we assume $-x\in {\cal V}_n$ where $-x$ is the vector satisfying $x+(-x)\in {\bf 0}$. Since such elements are not unique, this $-x$ is considered as a representative of the set of them.
\end{rem}

\subsection{Norm and Distance on ESDDs}

\begin{dfn}\label{dpr.3.1}  Let $x\in {\cal V}_m\subset {\cal V}$, $y\in {\cal V}_n\subset {\cal V}$, and ~$t=m\vee n$. Then the inner product of $x$ and $y$ is defined by
	\begin{align}\label{pr.3.1}
		\left<x\;,\;y\right>_{{\cal V}}:=\frac{1}{t}\left<x\otimes \J_{t/m}\;,\;y\otimes \J_{t/n}\right>,
	\end{align}
	where ~$\left<\cdot\;,\;\cdot\right>$ is the conventional inner product on $\R^t$.
	That is, if $x,~y\in \R^t$, then $\left<x,y\right>=\dsum_{i=1}^t x_iy_i$.
	The inner product defined by (\ref{pr.3.1}) is called the weighted inner product, because there is a weight coefficient $1/t$.
\end{dfn}

%

Using inner product, the norm of $x\in {\cal V}$ can be defined.

\begin{dfn}\label{dpr.3.3}
	The norm of $x\in {\cal V}$ is defined by
	\begin{align}\label{pr.3.6}
		\|x\|_{{\cal V}}:=\sqrt{\left<x\;,\;x\right>_{{\cal V}}}.
	\end{align}
\end{dfn}
%
One sees easily that $\|\cdot\|_{\cal V}$ defined by (\ref{pr.3.6}) satisfies linearity and triangle inequality, however, $\|x\|=0 \Rightarrow x\in \bf{0}$. It is also called a pseudo-norm.

Finally, we define the distance on ${\cal V}$.

\begin{dfn}\label{dpr.3.5}
	Let $x,~y\in {\cal V}$. The distance between $x$ and $y$ is defined by
	
	\begin{align}\label{pr.3.10}
		d_{{\cal V}}(x,y):=\|x\lvminus y\|_{{\cal V}}.
	\end{align}
\end{dfn}


It is easily verified that $d_{\cal V}$ satisfies symmetry and triangle inequality, however,
$d_{{\cal V}}(x,y)=0 \Rightarrow x\lvminus y\in \bf{0}$. Hence, this distance is called a pseudo-distance.

\begin{rem}\label{rpr.3.7}
	The distance defined on a vector space is, in general, required to be invariant under displacement. That is,
	\begin{align}\label{pr.3.14}
		d(x+z,y+z)=d(x,y), \quad x,y,z\in X.
	\end{align}
	It is easy to verify that the $d_{\cal V}$ defined by (\ref{pr.3.10}) satisfies (\ref{pr.3.14}).
\end{rem}

\subsection{Topology on ESDDs}

This subsection considers the topology on ${\cal V}$.
We refer to any standard textbook of topology for the basic topological concepts involved in this subsection, for instance, \cite{dug66,kel75}. In the following some topologies are considered.

\begin{itemize}
	\item Natural Topology:
\end{itemize}

Naturally, the topology on each $\R^n$ is considered as conventional topology. Precisely speaking, the open balls in $\R^n$ with center at $c=(c_1,c_2,\cdots,c_n)$, and radius $r>0$, are defined by
$$
\begin{array}{ccl}
	B^n_r(c):=\Big\{(x_1,\cdots,x_n)\in \R^n\;\Big|\sqrt{\dsum_{i=1}^n(x_i-c_i)^2}<r\Big\}.
\end{array}
$$
Taking
$$
B^n:=\{B^n_r(c)\;|\;c\in \R^n, r>0\}
$$
as topological basis, the topology on $\R^n$ generated by $B^n$ is the conventional topology on $\R^n$.

Then each $\R^n$, $n=1,2\cdots$, are considered as a set of clopen subsets in ${\cal V}$. Such a topology is called the natural topology on ${\cal V}$, denoted by ${\bf N}$.

The following properties are obvious.

\begin{prp}\label{ppr.6.1}
	\begin{itemize}
		\item[(i)] Assume ~$\emptyset\neq O_n\in \R^n$ is an open set, then it is also open in~$({\cal V},{\bf N})$.
		\item[(ii)] ~$({\cal V},{\bf N})$ is a second countable Hausdorff space.
	\end{itemize}
\end{prp}

\begin{itemize}
	\item Distance Topology:
\end{itemize}

Define open balls in ${\cal V}=\R^{\infty}$ by
$$
B_r(c):=\{x\in \R^{\infty}\;|\; d_{{\cal V}}(x,c)<r\},\quad c\in \R^{\infty}.
$$
Using $B:=\{B_r(c)\;|\;c\in \R^{\infty}, r>0\}$
as a topological basis, the topology generated by $B$ is called the distance topology on $\R^{\infty}$ deduced from $d_{\cal V}$, denoted by ${\bf D}$.

\begin{rem}\label{rpr.6.2}
	\begin{itemize}
		\item[(i)] Assume ~$\emptyset\neq O_n\in \R^n$ is an open set, it is not open under distance-deduced topology, i.e., not open in $({\cal V},{\bf D})$. This is because $\forall x\in O_n$ there exists a point $y=x\otimes \J_s\not\in O_n$. But $d(x,y)=0$, which means $x$ is not an interior point of $O_n$. Hence, $O_n$ is not open in $({\cal V},{\bf D})$.
		
		\item[(ii)] $({\cal V},{\bf D})$ is not a Hausdorff space.
		To see this, consider $x$ and $x\otimes \J_s$, $s>1$, which are two different points. But they are not separable in
		$({\cal V},{\bf D})$.
		It is clear that $({\cal V},{\bf D})$ is not even $T_0$.

		\item[(iii)] It is easy to see that if $O$ is open in $({\cal V},{\bf D})$, then $O$ is also open in $({\cal V},{\bf N})$.
		Hence ${\bf D}\subset {\bf N}$, that is, the distance-deduced topology ${\bf D}$ is rougher than the natural topology ${\bf N}$.
	\end{itemize}
\end{rem}
%
	%
	%
	%

\begin{itemize}
	\item  Product Topology:
\end{itemize}

One way to understand ${\cal V}=\R^{\infty}$ is to consider $\R^{\infty}=\prod_{n=1}^{\infty}\R^n$,
then the product topology is generated by the topological basis
$$
	B=\left\{\prod_{n=1}^{\infty}O_n\;\Bigg|\;O_n\subset \R^n~\mbox{is open, and}~O_n=\R^n ~\mbox{except for finite}~ n\right\}.
$$
The product topology is denoted by ${\bf P}$. It is easy to see that ${\bf P} = {\bf N}$.

\subsection{Equivalent vectors}

\begin{dfn}\label{dpr.2.1}
\begin{itemize}
\item[(i)] Let $x,~y\in {\cal V}$. $x$ and $y$ are said to be equivalent, denoted by $x\lra y$, if there exist two one-vectors $\J_{\a}$ and $\J_{\b}$, such that
\begin{align}\label{pr.2.1}
x\otimes \J_{\a}=y\otimes \J_{\b}.
\end{align}
\item[(ii)] The equivalence class of $x$ is denoted by $\bar{x} :=\left\{y\;\big|\;y\lra x\right\}$.
\end{itemize}
\end{dfn}

\begin{rem}\label{rpr.2.2}
Obviously $\lra$ is an equivalence relation. Assume $x,~y\in {\cal V}$. Then $x\lra y$, if and only if $x\lvminus y\in {\bf 0}$.
\end{rem}
For the equivalence we have the following properties.

\begin{thm}\label{tpr.2.4}
	\begin{itemize}
		\item[(i)] If $x\lra y$, then there exists a ~$\gamma\in {\cal V}$ such that
		\begin{align}\label{pr.2.3}
			x=\gamma\otimes \J_{\b},\quad y=\gamma\otimes \J_{\a}.
		\end{align}
		\item[(ii)] In each equivalence class $\bar{x}$ there exists unique smallest element $x_1\in \bar{x}$, such that $\bar{x}=\{x_1\otimes \J_k\;|\;k=1,2,\cdots\}$.
	\end{itemize}
\end{thm}

The proofs follow similarly as in \cite{che20}.

A partial order can be defined on $\cal V$.

\begin{dfn}\label{dpr.2.4}
A partial order, denoted by $\prec$, is defined as follows: Let $x,y\in \cal V$. If there exists a one-vector $\J_s$ such that ~$x\otimes \J_s=y$, then  $x\prec y$. For any equivalence class $\bar{x}$, $x_1\in \bar{x}$ is called the smallest element of $\bar{x}$, if $\forall y\in \bar{x}$, $y\prec x_1$ implies $y=x_1$.
\end{dfn}

\begin{rem}\label{rpr.2.5}
\begin{itemize}
\item[(i)]  If $x=y\otimes \J_s$, then $y$ is called a divisor vector of $x$, and $x$ is called a multiplier vector of $y$. This relation determines the order $y\prec x$.

\item[(ii)]  If (\ref{pr.2.3}) holds, and $\a,~\b$ are co-prime, then the $\gamma$ in Eq. (\ref{pr.2.3}) is called the maximum common divisor vector of $x$ and ~$y$, denoted by $\gamma=\gcd(x,y)$.

It is easy to prove that if $z$ is also a common divisor vector of $x$ and $y$, then $z\prec \gamma$. Moreover, the maximum common divisor vector is unique.
\item[(iii)]  If (\ref{pr.2.1}) holds and $\a,~\b$ are co-prime, then $\xi:=x\otimes \J_{\a}=y\otimes \J_{\b}$
is called the least common multiple vector of $x$ and $y$, denoted by $\xi=\lcm(x,y)$.
It is also easy to prove that  if $z$ is also a common multiple vector of $x$ and $y$, then $\xi \prec z$. Moreover, the least common multiple vector is also unique.


\end{itemize}
\end{rem}


\begin{prp}\label{ppr.2.6}
\begin{itemize}
\item[(i)] Assume $x\in {\cal V}$, then $(\bar{x},\prec)$ is a lattice \cite{che20}.
\item[(ii)] Assume $x,y\in {\cal V}$, then $(\bar{x},\prec)\approxeq (\bar{y},~\prec)$,
where $\approxeq$ stands for lattice isomorphism.  That is, any two equivalence classes as lattices are isomorphic.
\end{itemize}
\end{prp}

\noindent{\it Proof}. It is straightforward verifiable that $\forall u,~v\in \bar{x}$, $\sup(u,v)=\lcm(u,v)$; $\inf(u,v)=\gcd(u,v)$.
Then the first part is obvious. Assume
$\bar{x}=\{x_1,x_2,\cdots\}$ and $\bar{y}=\{y_1,y_2,\cdots\}$, where $x_i=x_1\otimes \J_i$, ~$i=1,2,\cdots$, etc. Define  $\pi:\bar{x}\ra \bar{y},~\pi(x_i)=y_i,~i=1,2,\cdots$,
Then one sees easily that $\pi$ is a lattice isomorphism.

\hfill $\Box$

The above arguments can be considered as special cases of that in Section 7 of \cite{che20} by choosing the vector multiplier as $\{\J_n\}$.

\section{Constructing DFESs From ESDDs}\label{s3}

\subsection{Quotient spaces as vector spaces}


\begin{dfn}\label{dqs.4.1}
The quotient space of $\cal V=\R^{\infty}$ under equivalence relation $\lra$ defined on it by (\ref{pr.2.1}), denoted by $\Omega$, is called the DFES. That is, 
\begin{align}\label{qs.1.1}
	\Omega:={\cal V}/\lra.
\end{align}
\begin{itemize}
\item[(i)] Let $\bar{x}\in \Omega$. The scalar product on $\Omega$ is defined by
\begin{align}\label{qs.1.4}
	a\bar{x}:=\overline{ax},\quad a\in \R.
\end{align}
\item[(ii)]
Let ~$\bar{x},~\bar{y}\in \Omega$. Then the addition of $\bar{x}$ and $\bar{y}$ is defined by
\begin{align}\label{qs.1.2}
\bar{x}\lvplus \bar{y}:=\overline{x\lvplus y}.
\end{align}
Correspondingly, the subtraction is defined by $\bar{x}\lvminus \bar{y}:=\bar{x}\lvplus (-\bar{y})$,
where $-\bar{y}:=\overline{-y}$.
\end{itemize}
\end{dfn}

It is easy to verify that the scalar product (\ref{qs.1.4}) and the addition (\ref{qs.1.2}) are well defined, that is, if $x\lra x'$ and $y\lra y'$, then $\overline{ax}\lra\overline{ax'}$, $\forall a\in \R$, and $x\lvplus y\lra x'\lvplus y'$. One may refer to \cite{che20} for proofs.



\begin{thm}\label{tqs.1.4} Using the addition defined as in (\ref{qs.1.2}) and the scalar product as in (\ref{qs.1.4}), $\Omega$ is a vector space.
\end{thm}

Consider the subspaces of ESDD and the corresponding subspaces of DFES.

\begin{dfn}\label{dqs.1.5}
\begin{itemize}
\item[(i)]  Let ~$p\in \Z_+$ be a positive integer. Define the $p$-upper truncated ESDD as
\begin{align*}
\R^{[p,\cdot]}:=\bigcup_{\{s\;\big|\; p|s\}}\R^s.
\end{align*}
\item[(ii)] Define ~$p$-upper truncated DFES as
\begin{align*}
\Omega^p:=\R^{[p,\cdot]}/\lra
~=\left\{\bar{x} \;\big|\; x_1\in \R^{pr},\;r\geq 1\right\}.
\end{align*}
\item[(iii)]  Define~$p$-lower truncated ESDD as
\begin{align}\label{qs.1.6}
\R^{[\cdot,p]}:=\bigcup_{\{s\;\big|\; s|p\}}\R^s.
\end{align}
\item[(iv)] Define~$p$-lower truncated DFES as
\begin{align*}
\Omega_p:=\R^{[\cdot,p]}/\lra
~=\left\{\bar{x} \;\big|\; x_1\in \R^{s},\;s|p\right\}.
\end{align*}
\end{itemize}
\end{dfn}

The next proposition is an immediate consequence of the definition.

\begin{prp}\label{pqs.1.6}
\begin{itemize}
\item[(i)] ~$\Omega^p$, and ~$\Omega_p$, ~$p=1,2,\cdots$ are subspaces of ~$\Omega$;
\item[(ii)] If ~$i|j$, Then, ~$\Omega^j$ is a subspace of ~$\Omega^i$,
$\Omega_i$ is a subspace of $\Omega_j$.
\end{itemize}
\end{prp}

%

%

\begin{exa}\label{eqs.1.603} The lattice structure on $\R^{\infty}$ can be transferred to $\Omega$:
\begin{itemize}
\item[(i)] Define $\Omega_{(n)}:=\R^n/\lra,\quad n=1,2,\cdots$.
Then $\Omega=\bigcup_{n=1}^{\infty}\Omega_{(n)}$.  Define $\Omega_{(m)}\prec \Omega_{(n)}\Leftrightarrow \R^m\prec \R^n$,
then it is obvious that $(\Omega,\prec)$ is a lattice with $\sup(\Omega_{(p)},\Omega_{(q)})=\Omega_{(p\vee q)}$, $\inf(\Omega_{(p)},\Omega_{(q)})=\Omega_{(p\wedge q)}$.
\item[(ii)] $\Omega_{p}$ is an ideal of $\Omega$.

\item[(iii)] $\Omega^{p}$ is a filter of $\Omega$.
\end{itemize}
\end{exa}

In fact, $\Omega$ has the same lattice structure as its filters.

\begin{prp}\label{pqs.1.7}
Let ~$p>1$. The filter $\Omega^p$ is lattice isomorphic to~$\Omega$.
\end{prp}

\noindent{\it Proof}. 
Define a mapping~$\varphi:\Omega^p\ra \Omega,~\varphi(\Omega_{(np)}):=\Omega_{(p)}$.
Then it is easy to verify that $\varphi$ is a lattice isomorphism.
\hfill $\Box$

\subsection{Topology on DFESs}

First, we extend the inner product over ESDD $\R^{\infty}$ to DFES $\Omega$.

\begin{dfn}\label{dqs.2.1} Let ~$\bar{x}, ~\bar{y}\in \Omega$. Define their inner product as
\begin{align}\label{qs.2.1}
\left<\bar{x}\;,\;\bar{y}\right>_{{\cal V}}:=\left<x\;,\;y\right>_{{\cal V}}, \quad x\in \bar{x}, ~~y\in \bar{y},
\end{align}
where $\langle \cdot,\cdot\rangle_{\cal V}$ is defined as in (\ref{pr.3.1}).
\end{dfn}

The following proposition shows Definition \ref{dqs.2.1} is well defined.

\begin{prp}\label{pqs.2.2} (\ref{qs.2.1}) is properly defined. That is, it is independent of the choice of representatives $x$ and $y$.
\end{prp}
\noindent{\it Proof}. Assume $x_1\lra x_2$ and $y_1\lra y_2$. According to Theorem \ref{tpr.2.4},
there exist $x_0\in \R^s$ and $y_0\in \R^t$,
such that
$$
\begin{array}{l}
x_1=x_0\otimes \J_{\a}; \quad x_2=x_0\otimes \J_{\b}, \\
y_1=y_0\otimes \J_{p}; \quad y_2=y_0\otimes \J_{q}. \\
\end{array}
$$
First, we prove two facts:
\begin{itemize}
\item Fact 1:
Let $s\wedge t=\xi$, and $s=a\xi$, $t=b\xi$, where $\a\wedge b=1$. If $f,g$ satisfy $sf=tg$,
then, $af\xi=bg\xi$, i.e., $af=bg$. Since $a\wedge b=1$, there exists a $c$ such that $f=cb,\quad g=ca$.
\item Fact 2: $\left<x,y\right>_{{\cal V}}=\left<x\otimes \J_s,y\otimes \J_s\right>_{{\cal V}}$.
\end{itemize}
These facts can be verified by definition directly.

Next, we consider
$$
\begin{array}{ccl}
\left<x_1,y_1\right>_{{\cal V}}&=&\left<x_0\otimes \J_{\a},y_0\otimes \J_{p}\right>_{{\cal V}}\\
~&=&\left<x_0\otimes \J_{\a}\otimes \J_{\frac{s\a\vee tp}{s\a}},y_0\otimes \J_{p}\otimes \J_{\frac{s\a\vee tp}{tp}}\right>_{{\cal V}}\\
~&=&\left<x_0\otimes \J_{\frac{s\a\vee tp}{s}},y_0\otimes \J_{\frac{s\a\vee tp}{t}}\right>_{{\cal V}}.
\end{array}
$$
Hence we have $s \frac{s\a\vee tp}{s}=t \frac{s\a\vee tp}{t}$.
Using Fact 1, one sees that $\frac{s\a\vee tp}{s}=cb;~\frac{s\a\vee tp}{t}=ca$.
Using Fact 2 yields
$$
\left<x_1,y_1\right>_{{\cal V}}=\left<x_0\otimes \J_{cb}, y_0\otimes \J_{ca}\right>_{{\cal V}}
= \left<x_0\otimes \J_{b}, y_0\otimes \J_{a}\right>_{{\cal V}}.
$$
Similarly, we have $\left<x_2,y_2\right>_{{\cal V}}= \left<x_0\otimes \J_{b}, y_0\otimes \J_{a}\right>_{{\cal V}}$.
The conclusion follows.
\hfill $\Box$

Since $\Omega$ is a vector space, (\ref{qs.2.1}) defines an inner product on $\Omega$. This inner product has the following properties.

\begin{prp}\label{pqs.2.3} $\Omega$ with the inner product defined by (\ref{qs.2.1}) is an inner product space. But it is not a Hilbert space.
\end{prp}

\noindent{\it Proof}.  Obviously $\Omega$ is an inner product space. To see that it is not a Hilbert space, we construct a sequence as follows:
$$
x_1=a\in \R;~
x_{i+1}=x_i\otimes \J_2+\frac{1}{2^{i+1}}\left(\d_{2^{i+1}}^1-\d_{2^{i+1}}^2\right),\quad i=1,2,\cdots.
$$
It is obvious that this sequence is a Cauchy sequence. But it does not converge to any point  ~$x\in {\cal V}$. Let ~$\bar{x}_i:=\overline{x_i}$. According to Proposition \ref{pqs.2.2}, it is easy to see that ~$\{\bar{x}_i\}$ is also a Cauchy sequence in $\Omega$, but it can not converge to any point in $\Omega$.
\hfill $\Box$

$\forall x\in \R^{\infty}$, $\varphi_x:y\mapsto \left<x\;,\; y\right>_{{\cal V}}$ gives a mapping $\varphi_x:~\R^{\infty}\ra \R$.
Similarly, a point $\bar{x} \in \Omega$ can be used to construct a mapping $\varphi_{\bar{x}}:~\Omega \ra \R, ~\bar{y}\mapsto \left<\bar{x}\;,\; \bar{y}\right>_{{\cal V}}$.
Conversely, not every linear mapping  ~$\varphi: \Sigma\ra \R$ can be expressed as a mapping deduced by an element as $\varphi_{\bar{x}}$. This is because $\Omega$ is an infinite dimensional vector space, while each element $\bar{x}\in \Omega$ is a finite dimensional element.

Using the inner product defined by (\ref{qs.2.1}), the norm and distance on $\Omega$ are also well defined.

\begin{dfn}\label{dqs.2.4}
\begin{itemize}
\item[(i)] Let $\bar{x}\in \Omega$. The norm of $\bar{x}$ is defined by
\begin{align}\label{qs.2.4}
\|\bar{x}\|_{{\cal V}}:=\|x\|_{{\cal V}}.
\end{align}
\item[(ii)] Let $\bar{x},~\bar{y}\in \Omega$. The distance between $\bar{x}$ and $\bar{y}$ is defined as
\begin{align}\label{qs.2.5}
d_{{\cal V}}(\bar{x},\bar{y}):=d_{{\cal V}}(x,y).
\end{align}
\end{itemize}
\end{dfn}

According to Proposition \ref{pqs.2.2},  (\ref{qs.2.4})(\ref{qs.2.5}) are both well defined.

Finally, As a topological space, the topology on $\Omega$ is deduced by the distance. This topology is equivalent to the quotient topology of $\left(\R^{\infty},{\bf N}\right)$ over equivalence. That is, the glued topology inherited from $\left(\R^{\infty},{\bf N}\right)$.

As a topological space, $\Omega$ has the following properties.

\begin{prp}\label{pqs.2.5}
{$\Omega$ is a second countable Hausdorff space.}
\end{prp}

\noindent{\it Proof}. 
Since $\R^n$ is second countable, denote by ~$\{O^n_i\;|\;i=1,2,\cdots\}$ its countable topological bases. Then
$\bigcup_{n=1}^{\infty}\bigcup_{i=1}^{\infty}O^n_i$ is a topological basis of $\R^{\infty}$, which is also countable. Hence, as its quotient space, $\Omega={\cal V}/\lra$ is also second countable.

Since $\Omega$ is a metric space, then ~$\bar{x}\neq \bar{y}$, if and only if, $d_{{\cal V}}(\bar{x},\bar{y})>0$. It is obvious that this space is a Hausdorff space. (In fact, it is easy to see that this space is $T_4$.)
\hfill $\Box$

\begin{dfn}\label{dqs.4.0} Let ~$\bar{x}\in \Omega$. The dimension of $\bar{x}$, denoted by ~$\dim(\bar{x})$, is the dimension of the smallest element in $\bar{x}$. That is, $\dim(\bar{x})=\dim(x_1)=\min_{x\in \bar{x}}\dim(x)$.
\end{dfn}

\begin{rem}\label{rqs.2.6}
\begin{itemize}
\item[(i)] Note that $\bar{x}=\{x_1\otimes \J_n\;|\;n=1,2,\cdots\}$,
it is clear that $\bar{x}$ can be considered as $x_1$ and the images of merging $x_1\in \R^s$ into
$\R^{ns}$, $n=2,3,\cdots$. Hence $x_1$ is the essential element in $\bar{x}$, which determines $\bar{x}$ completely. This fact shows that the Definition \ref{dqs.4.0} is reasonable.
\item[(ii)] It is surprising that $\Omega$ is a topological vector space with each point $\bar{x}\in \Omega$ having its own dimension $1\leq \dim(\bar{x})<\infty$. Hence, the  {\bf DFES (dimension-free Euclidean space) is a totally new mathematical object}.
\end{itemize}
\end{rem}

\subsection{Fiber bundle structure on ESDDs and DFESs}
First, we recall the definition of a fibre bundle.

\begin{dfn}\label{dqs.3.1}\cite{hus94} Let $T$ and $B$ be two topological spaces, $\Pr:T\ra B$
is a continuous surjective mapping. Then $(T,\Pr,B)$
is called a fiber bundle, where $T$ is the total space, $B$ is the base space. $\forall b\in B$, ${\Pr}^{-1}(b)$ is called the fiber at $b$.
\end{dfn}

The following result comes from the  definition immediately.

\begin{prp}\label{pqs.3.2} Let $T=(\cal V,{\bf N})$ be the total space, $B=(\Omega,{\bf D})$ be the base space, and $\Pr: T\ra B$ be the natural projection, i.e., ~$x \mapsto \bar{x}$. Then $(\cal V, {\bf N}) \stackrel{\Pr}{\longrightarrow} (\Omega,{\bf D})$
is a fiber bundle, which is called the dimension-free Euclidean bundle (DFEB).
\end{prp}

The DFEB is said to be a discrete bundle, because the bundle at each point $\bar{x}$ is a discrete countable (topological) subspace of ESDD $\R^{\infty}$.

\begin{dfn}\label{dqs.3.3}
\begin{itemize}
\item[(i)] Two fiber bundles $(T_i, \Pr_i, B_i)$, ~$i=1,2$ are called homomorphic, if there exist two continuous mappings $\pi:T_1\ra T_2$ and $\varphi:B_1\ra B_2$, such that the diagram (\ref{d1}) is commutative. In addition, if both $\pi$ and $\varphi$ are bijective, and $\pi^{-1}:T_2\ra T_1$ and $\varphi^{-1}:B_2\ra B_1$ are also making (\ref{d1}) commutative, $(T_i, Pr_i, B_i)$, ~$i=1,2$ are said to be isomorphic.
	\begin{align}\label{d1}
	\xymatrix{
		T_1\ar[r]^{\pi}\ar[d]^{\mathrm{Pr_1}}&T_2\ar[d]^{\mathrm{Pr_2}}\\
		B_1\ar[r]^{\varphi}&B_2
	}
	\end{align}
\item[(ii)] Two fiber bundles on $B$, denoted by $(T_i, \Pr_i, B)$, ~$i=1,2$, are called homomorphic, if there exists a continuous mapping $\pi:T_1\ra T_2$, such that the diagram (\ref{d2}) is commutative. In addition, if $\pi$ is bijective, and $\pi^{-1}:T_2\ra T_1$ making (\ref{d2}) commutative, $(T_i, \Pr_i, B)$, ~$i=1,2$ are said to be isomorphic.
	\begin{align}\label{d2}
		\xymatrix{
			T_1\ar[r]^{\pi}\ar[d]^{\mathrm{Pr_1}}&T_2\ar[dl]^{\mathrm{Pr_2}}\\
			B_1
		}
	\end{align}
\end{itemize}
\end{dfn}

%
%

%

\begin{exa}\label{eqs.3.4} Consider~$\left(\R^{[p,\cdot]},\Pr,\Omega\right)$ and
$\left(\R^{\infty},\Pr,\Omega\right)$.
Define
$\pi: \R^{[p,\cdot]}\hookrightarrow \R^{\infty}$ as the including mapping. Then it is obvious that $\pi$ is a fiber bundle homomorphism, since the following diagram commutes.
	\begin{align*}
	\xymatrix{
		\R^{[p,\cdot]}\ar[r]^{\pi}\ar[d]^{\mathrm{Pr}}&\R^{\infty}\ar[dl]^{\mathrm{Pr}}\\
		\Omega
	}
	\end{align*}
%
%
%

\end{exa}

\subsection{Coordinate neighbourhoods}

To establish a differential structure on DFES, we need a ``local coordinate neighborhood" for each point $\bar{x}\in \Omega$. Since $\Omega$ is a dimension-free space, the coordinate neighborhoods are not classical ones in standard differential manifold. In fact, they are sub-bundles of DFEB.


\begin{dfn}\label{dqs.4.2} Let $\bar{x}\in \Omega$, and ~$\dim(\bar{x})=p$. Assume ~$O_{\bar{x}}$ is an open neighborhood of $\bar{x}\in \Omega$. That is, ~$\bar{x}\in O_{\bar{x}}$, and ~$O_{\bar{x}}\subset \Omega$ is open.
Then
\begin{align*}
{\cal V}_{O_{\bar{x}}}:={\Pr}^{-1}\left(O_{\bar{x}}\right)\bigcap \R^{[p,\cdot]}
\end{align*}
with $\R^{[p,\cdot]}$ defined as in (\ref{qs.1.6}), is called the set of coordinate charts of ~$\bar{x}$, $({\cal V}_{O_{\bar{x}}},\Pr, O_{\bar{x}})$
is called the bundle of coordinate neighborhood of $\bar{x}$.
\begin{align*}
{\cal V}^r_{O_{\bar{x}}}:={\Pr}^{-1}\left(O_{\bar{x}}\right)\bigcap \R^{rp},\quad r=1,2,\cdots
\end{align*}
is called a leaf of the bundle of coordinate neighborhood bundle of $\bar{x}$.
\end{dfn}

An example is given in the following to depict the bundle of coordinate neighborhood.

\begin{exa}\label{eqs.4.3} Assume $x=(\a,\a,\b,\b)^T\in\R^4$, then $\bar{x}=\{x_1,x_2,\cdots\}$,
where, $x_1=(\a,\b)\in \R^2$. Hence $\dim(\bar{x})=2$. Consider $O_{\bar{x}}=B_r(\bar{x})\subset \Omega$, which is an open ball neighborhood of $\bar{x}$. Then the set of coordinate charts,  deduced by $O_{\bar{x}}$, is ${\cal V}_O=\left\{B_{r_1}(x_1),B_{r_2}(x_2),\cdots\right\}$,
where  $r_k=1/\sqrt{2k}$, $x_k=(\a,\b)^T\otimes \J_k$,  $k=1,2,\cdots$.
The bundle of coordinate neighborhood of $\bar{x}$ is
$
({\cal V}_{O_{\bar{x}}},{\Pr},O_{\bar{x}}).
$

Fig. \ref{Figqs.4.1} demonstrates the bundle of coordinate neighborhood of $\bar{x}$.

\begin{figure}
\centering
\setlength{\unitlength}{6mm}
\begin{picture}(10,11)\thicklines
\put(1,2){\line(1,0){6}}
\put(3,4){\line(1,0){6}}
\put(1,2){\line(1,1){2}}
\put(7,2){\line(1,1){2}}
\put(5,3){\oval(3.8,1.8)}
\put(5,3){\vector(1,0){1.9}}
\put(5.5,3.2){$r_1$}
\put(7.5,3.4){$\R^2$}
\put(1,5){\line(1,0){6}}
\put(3,7){\line(1,0){6}}
\put(1,5){\line(1,1){2}}
\put(7,5){\line(1,1){2}}
\put(5,6){\oval(3,1.5)}
\put(5,6){\vector(1,0){1.5}}
\put(5.2,6.2){$r_2$}
\put(7.5,6.4){$\R^4$}
\put(1,8){\line(1,0){6}}
\put(3,10){\line(1,0){6}}
\put(1,8){\line(1,1){2}}
\put(7,8){\line(1,1){2}}
\put(5,9){\oval(2.4,1.2)}
\put(5,9){\vector(1,0){1.2}}
\put(5.2,9.2){$r_3$}
\put(7.5,9.4){$\R^6$}
\put(5,11){\vector(0,-1){9.7}}
\put(5.2,10.5){$\Pr$}
\put(4.3,3){$x_1$}
\put(4.3,6){$x_2$}
\put(4.3,9){$x_3$}
\put(4.8,0.5){$\bar{x}\in O_{\bar{x}}\subset \Omega$}
\end{picture}
\caption{Bundle of coordinate neighborhood\label{Figqs.4.1}}
\end{figure}
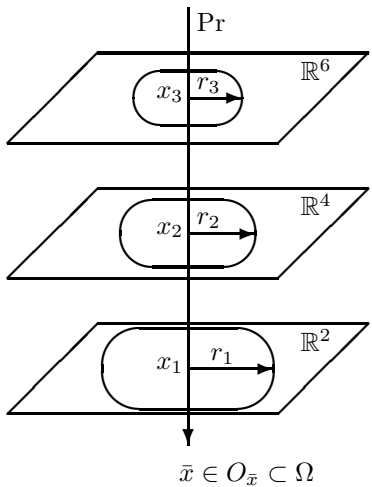

\end{exa}

Note that the set of coordinate charts ${\cal V}_O$ does not include all the inverse image of  ~$O$, i.e., $
{\cal V}_O \subsetneqq {\Pr}^{-1}(O)$.
But it can provide coordinates for all points within $O$. The following proposition shows this.

\begin{prp}\label{pqs.4.4} Assume $\bar{y}\in O$, then ${\Pr}^{-1}(\bar{y})\bigcap {\cal V}_O\neq \emptyset$.
\end{prp}
\noindent{\it Proof}.  Assume ~$\bar{y}\in O$, ~$\dim(\bar{x})=p$, ~$\dim(\bar{y})=q$, $r=p\vee q$, then $y_{r/q}\in {\Pr}^{-1}(O)\bigcap \R^r\subset {\cal V}_O$.
\hfill $\Box$

\begin{rem}\label{rqs.4.401} Assume $\bar{x}\in \Omega$ with $\dim(\bar{x})=p$. Then ${\cal V}_{\Omega_{\bar{x}}}:={\Pr}^{-1}(\Omega)\bigcap \R^{[p,\cdot]}=\R^{[p,\cdot]}$
is a coordinate neighborhood of $\bar{x}$, which is the largest  coordinate neighborhood of $\bar{x}$.
When the DFES is considered, we can simply use this coordinate neighborhood, then the corresponding coordinates are called the global coordinates. The general definition is mainly for DFM.
\end{rem}

\section{ Projections on ESDDs and Continuous Functions on DFESs}\label{s4}

In this section we introduce the cross-dimensional projections on ESDD, which is crucial for prolonging functions on a finite-dimensional Euclidean space to construct continuous functions on DFES.

\subsection{Cross-dimensional projection of a vector}

\begin{dfn}\label{dcg.1.1} Assume  $\xi\in {\cal V}_n$. A cross-dimensional projection of $\xi$ to ${\cal V}_m$, denoted by $\pi^n_m(\xi)$, is defined as follows:
	\begin{align}\label{cg.1.1}
		\pi^n_m(\xi):=\underset{{x\in {\cal V}_m}}{\mathrm{argmin}}\|\xi-x\|_{{\cal V}}.
	\end{align}
\end{dfn}

Assume $t=\lcm(n,m)=t$ and denote ~$\a:=t/n$, ~$\b:=t/m$.  Then the distance between $\xi$ and $x\in {\cal V}$ is $\D:=\|\xi-x\|^2_{{\cal V}}=\frac{1}{t}\|\xi\otimes \J_{\a}-x\otimes \J_{\b}\|^2$.

Denote $\xi\otimes \J_{\a}:=(\eta_1,\eta_2,\cdots,\eta_t)^T$,
where $\eta_j=\xi_i$, $(i-1)\a+1\leq j\leq i\a;\; i=1,\cdots,n$.
Then, $\D=\frac{1}{t}\dsum_{i=1}^m\dsum_{j=1}^{\b}\left(\eta_{(i-1)\b+j}-x_i\right)^2$.
Setting $\frac{\pa \D}{\pa x_i}=0,~i=1,\cdots, m$ yields
\begin{align}\label{cg.1.3}
	x_i=\frac{1}{m}\left(\dsum_{j=1}^{\b} \eta_{(i-1)\b+j}\right),\quad i=1,\cdots,m.
\end{align}
That is, ~$\pi^n_m(\xi)=x$. Moreover, it is easy to verify the following orthogonality, i.e. $\left<\xi\lvminus x, x\right>_{{\cal V}}=0$.
The above argument leads to the following conclusion:

\begin{prp} \label{pcg.1.2} Let $\xi\in {\cal V}_n$. Then the projection of $\xi$ on ~${\cal V}_m$, say, $x$, can be calculated by (\ref{cg.1.3}). Moreover, $\xi\lvminus x$  and $x$ are orthogonal.
\end{prp}

\begin{exa}\label{ecg.1.3} Assume $\xi=[1,0,-1,0,1,2,-2]^T\in \R^7$. Consider its projection on $\R^3$, denoted by $\pi^7_3(\xi):=x$.
	Then $\eta=\xi\otimes \J_3$.  Denote by $x=[x_1,x_2,x_3]^T$, then
	$$
		x_1 =\frac{1}{7}\dsum_{j=1}^7\eta_j=0.2857,\quad
		x_2 =\frac{1}{7}\dsum_{j=8}^{14}\eta_j=0,\quad
		x_3 =\frac{1}{7}\dsum_{j=15}^{21}\eta_j=0.1429.
	$$
	Moreover,
	$$
	\begin{array}{rcl}
		\xi\lvminus x&=&[0.7143,0.7143,0.7143,-0.2857,-0.2857,-0.2857,-1.2857,-1,-1,\\
		~&~&0,0,0,1,1,0.8571,1.8571,1.8571,1.8571,-2.1429,-2.1429,-2.1429].
	\end{array}
	$$
	
\end{exa}

Since the projection of a vector to a space of different dimension $\pi^n_m$ is a linear mapping, it can be expressed by a matrix. Assume there exists a matrix $\Pi^n_m$, such that the projection of $\xi\in \R^n$ to $\R^m$ can be expressed as
\begin{align}\label{cg.2.1}
	\pi^n_m(\xi)=\Pi^n_m\xi,\quad \xi\in {\cal V}_n.
\end{align}
We try to figure out this matrix.

Let $\lcm(n,m)=t$, $\a:=t/n$, and $\b:=t/m$. Then
$$
	\eta=\xi\otimes \J_{\a}=\left(I_n\otimes \J_{\a}\right)\xi,\quad
	x=\frac{1}{\b}\left(I_m\otimes \J_{\b}^T\right)\eta
	=\frac{1}{\b}\left(I_m\otimes \J_{\b}^T\right)\left(I_n\otimes \J_{\a}\right)\xi.
$$
Hence we have
\begin{align}\label{cg.2.2}
	\Pi^n_m=\frac{1}{\b}\left(I_m\otimes \J_{\b}^T\right)\left(I_n\otimes \J_{\a}\right).
\end{align}

Using this structure, we have the following result.

\begin{lem}\label{lcg.2.1}
	\begin{itemize}
		\item[(i)] Let $n \geq m$. Then $\Pi^n_m$ is of full row rank. Hence, $\Pi^n_m(\Pi^n_m)^T$ is invertible.
		\item[(ii)] Let $n \leq m$. Then $\Pi^n_m$ is of full column rank. Hence, $(\Pi^n_m)^T\Pi^n_m$ is invertible.
	\end{itemize}
\end{lem}
\noindent{\it Proof}. 
	\begin{itemize}
		\item[(i)] Assume $n\geq m$.  When $n=m$, $\Pi^n_m (\Pi^n_m)^T$ is an identity matrix, the conclusion is trivial. We, therefore, need only to consider the case when $n>m$. According to the structure of $\Pi^n_m$ determined by (\ref{cg.2.2}), it is easy to see that each row of $\Pi^n_m$ contains at least two non-zero elements. Moreover, when $j>i$ the column of non-zero element in row $i$ is prior to the column of non-zero element in row $j$, and only when $j=i+1$ there is an overlapped column. This structure ensures the full row rank of $\Pi^n_m$. Hence, $\Pi^n_m (\Pi^n_m)^T$ is invertible.
		\item[(ii)] According to  (\ref{cg.2.2}), one sees  easily that $\Pi^m_n=\frac{\b}{\a}\left(\Pi^n_m\right)^T$, hence, the full column rank of $\Pi^n_m$ comes from the full row rank of $\Pi^m_n$.
	\end{itemize}
\hfill $\Box$

The following proposition shows that the projection from factor dimension space to multiple dimension space does not lose information.

\begin{prp}\label{pcg.1.4} Let $X\in \R^m$. Project it to $\R^{km}$ and then project the image back to $\R^m$, the vector $X$ remains unchanged. That is,
	\begin{align}\label{cg.1.4}
		\Pi^{km}_m\Pi^m_{km}=I_m.
	\end{align}
\end{prp}
\noindent{\it Proof}. $\Pi^{km}_m\Pi^m_{km}=\frac{1}{k}\left(I_m\otimes \J_k^T\right)
		\left(I_m\otimes \J_k\right)=\frac{1}{k}\left(I_m\otimes \J_k^T\J_k\right)=I_m$.
\hfill $\Box$

\subsection{Continuous functions on DFESs}

Now we are ready to define continuous functions on $\Omega$.

\begin{dfn}\label{pqs.4.5} Let $f:\Omega\ra \R$ be a real function on $\Omega$.
	\begin{itemize}
		\item[(i)] Define $f:\R^{\infty}\ra \R$, $x\mapsto f(\bar{x})$.
		\item[(ii)] If for each point $\bar{x}\in \Omega$ there exists a neighborhood $O_{\bar{x}}$ of ${\bar{x}}$ such that on each leaf ${\cal V}^r_{O_{\bar{x}}}\subset \R^{rp}$ $f\in C({\cal V}^r_{O_{\bar{x}}})$, then $f$ is called a continuous function on $\Omega$.
		\item[(iii)] If on each leaf of the bundle of coordinate neighborhood $f\in C^r({\cal V}^r_{O_{\bar{x}}})$, then $f$ is called a $C^r$ function on $\Omega$, where $r=1,2,\cdots,\infty,\omega$, $r=\omega$ means $f$ is an analytic function.
	\end{itemize}
\end{dfn}

\begin{rem}\label{rqs.4.6} In definition \ref{dqs.4.2}, the set of coordinate neighborhood is used. In fact, up to now only global coordinates are used. So the definition can also use global coordinates. That is, consider $\R^{rp}$ as each leaf of the bundle of coordinate neighborhood.
\end{rem}

Constructing a differentiable function on $\Omega$ directly is very difficult. Our technique to construct such a function is to transfer a smooth function on $\R^{\infty}$ to $\Omega$. Note that $\R^n$ is a clopen subset of $\R^{\infty}$. $f:\R^{\infty}\ra \R$ is continuous, if and only if, $f_n:=f|_{\R^n}$, $n=1,2,\cdots$, are continuous. Hence, it is reasonable to transfer an $f\in C^r(\R^n)$ to $\Omega$.

\begin{dfn}\label{rqs.4.7} Let $f\in C^r(\R^n)$. Define $\bar{f}:\Omega\ra \R$ as follows:
	Let $\bar{x}\in \Omega$ and $\dim(\bar{x})=m$. Then
	\begin{align}\label{qs.4.5}
		\bar{f}(\bar{x}):=f(\Pi^m_n(x_1)),\quad \bar{x}\in \Omega,
	\end{align}
	where $x_1\in \bar{x}$ is the smallest element in $\bar{x}$.
\end{dfn}

\begin{prp}\label{pqs.4.8} Assume $f\in C^r(\R^n)$, then the function $\bar{f}$ defined by (\ref{qs.4.5}) is $C^r$, that is, $\bar{f}\in C^r(\Omega)$.
\end{prp}
\noindent{\it Proof}.  Given $\bar{x}\in \Omega$, where $\dim(\bar{x})=m$. Consider a leaf of the bundle of coordinate neighborhood ${\cal V}^r_{O_{\bar{x}}}$ of $\bar{x}$. Assume $y\in {\cal V}^r_{O_{\bar{x}}}$, consider the following two cases:
	\begin{itemize}
		\item Case $1$: ~$y\in \R^{rm}$ is the smallest element of $\bar{y}$.
		By definition,
		\begin{align}\label{qs.4.6}
			\bar{f}(y)=f(\Pi^{rm}_m y).
		\end{align}
		\item Case $2$: $y_1\in \bar{y}$ is the smallest element of $\bar{y}$ and $\dim(y_1)=\xi$. Then there exists $s$ such that $y=y_1\otimes \J_s$. Since $y\in \R^{rm}$, then $\xi s=mr$.
		By definition,
		\begin{align}\label{qs.4.7}
			\bar{f}(y)=f(\bar{y})=f(\Pi^{\xi}_m y_1).
		\end{align}
		Denote $z_0:=\Pi^{\xi}_m y_1\in \R^m$. 
		Then $z_0$ is the point on $\R^m$, which is closest to $y_1$. Since $y\lra y_1$, According to Proposition \ref{pqs.2.2}, we know $d_{{\cal V}}(z,y)=d_{{\cal V}}(z,y_1),~z\in \R^m$.
		Hence, $z_0$ is also the point on $\R^m$ which is closest to $y$. That is, $\Pi^{mr}_m y=z_0=\Pi^{\xi}_m y_1$.
		Hence, (\ref{qs.4.7}) becomes (\ref{qs.4.6}). It is obvious that $\bar{f}$ is a $C^r$ function on ${\cal V}^r_{O_{\bar{x}}}$.
	\end{itemize}
\hfill $\Box$

The following is a simple example.

\begin{exa}\label{eqs.4.9} Given
	\begin{align}\label{qs.4.8}
		f(x_1,x_2,x_3)=x_1+x^2_2-x_3\in C^{\omega}(\R^3).
	\end{align}
	\begin{itemize}
		\item[(i)] Assume $\bar{y}\in \Omega$, where $y_1=(\xi_1,\xi_2,\xi_3,\xi_4,\xi_5)^T\in \R^5$.
		It is easy to calculate that
		$$
		\Pi^5_3=\frac{1}{5}\left(I_3\otimes \J_5^T\right)\left(I_5\otimes \J_3\right)=\frac{1}{5}\begin{bmatrix}
							3&2&0&0&0\\
							0&1&3&1&0\\
							0&0&0&2&3\\
						\end{bmatrix}.
		$$
		Hence we have
		$$
			\bar{f}(\bar{y})=f(\Pi^5_3y_1)=\frac{1}{5}(3\xi_1+2\xi_2)+\frac{1}{25}(\xi_2+3\xi_3+\xi_4)^2-\frac{1}{5}(2\xi_4+3\xi_5).
		$$
		\item[(ii)] Assume $\bar{y}\in \Omega$, where $y_1=(\xi_1,\xi_2)\in \R^2$.
		
		Consider ~${\cal V}^1_O$: since $\Pi^2_3=\frac{1}{2}\begin{bmatrix}
			1&0\\
			0.5&0.5\\
			0&1
		\end{bmatrix}$, we have $\bar{f}|_{{\cal V}^1_O}=\xi_1+\frac{1}{4}(\xi_1+\xi_2)^2-\xi_2$.
		
		Consider ${\cal V}^2_O$: since $\Pi^4_3=\frac{1}{4}\begin{bmatrix}
			3&1&0&0\\
			0&2&2&0\\
			0&0&1&2
		\end{bmatrix}$, we have $\bar{f}|_{{\cal V}^2_O}=\frac{1}{4}(3\xi_1+\xi_2)+\frac{1}{16}(\xi_2+\xi_3)^2-\frac{1}{4}(\xi_3+3\xi_4)$.
	\end{itemize}
\end{exa}

\subsection{Least square approximation of linear systems}

Consider a linear system
\begin{align}\label{cg.2.3}
	\xi(t+1)=A\xi(t),\quad \xi(t)\in \R^n.
\end{align}

Our goal is to find a matrix $A_{\pi}\in {\cal M}_{m\times m}$, and  construct a linear system on
$\R^m$ as
\begin{align}\label{cg.2.4}
	x(t+1)=A_{\pi}x(t),\quad x(t)\in \R^m.
\end{align}
Then take (\ref{cg.2.4}) as the projected system of (\ref{cg.2.3}) on $\R^m$.

We are mainly concerning about the trajectories. The trajectory of the idea projected system should satisfy the same projection relation. That is,
\begin{align}\label{cg.2.5}
	x(t,\pi(\xi_0))=\pi^n_m(\xi(t,\xi_0)).
\end{align}
Unfortunately, it is, in general, impossible to realize this.  So we can only search such a system that makes the error of (\ref{cg.2.5}) smallest. Then a practical way is that we can search for the least square approximation.

\begin{prp}\label{pcg.2.1} Let system (\ref{cg.2.3}) be approximated by system (\ref{cg.2.4}). Then the least square approximation satisfies
	\begin{align}\label{cg.2.6}
		A_{\pi}=\begin{cases}
			\Pi^n_mA(\Pi^n_m)^T\left(\Pi^n_m (\Pi^n_m)^T\right)^{-1}\quad n\geq m\\
			\Pi^n_mA\left((\Pi^n_m)^T\Pi^n_m\right)^{-1}(\Pi^n_m)^T\quad n< m.
		\end{cases}
	\end{align}
\end{prp}
\noindent{\it Proof}. 
	From (\ref{cg.2.5}) we have $x(t)=\Pi^n_m\xi(t),~\mbox{with}~x_0=\Pi^n_m\xi_0$.
	Plugging it into  (\ref{cg.2.4}) yields
	\begin{align}\label{cg.2.7}
		\Pi^n_m\xi(t+1)=A_{\pi}\Pi^n_m\xi(t).
	\end{align}
	Using (\ref{cg.2.3}) and noting that $\xi(t)$ is arbitrary, we have
	\begin{align}\label{cg.2.8}
		\Pi^n_mA=A_{\pi}\Pi^n_m.
	\end{align}
	
	Assume ~$n\geq m$, right multiplying both sides of  (\ref{cg.2.8}) by $(\Pi^n_m)^T\left(\Pi^n_m (\Pi^n_m)^T\right)^{-1}$ yields the first equality of  (\ref{cg.2.6}).
	
	Assume $n<m$, we search a solution of the following form: $A_{\pi}=\tilde{A}(\Pi^n_m)^T$.
	Then the least square solution $\tilde{A}$ can be obtained as $\tilde{A}=\Pi^n_m A\left((\Pi^n_m)^T\Pi^n_m\right)^{-1}$.
	Hence, we have $A_{\pi}=\Pi^n_mA\left((\Pi^n_m)^T\Pi^n_m\right)^{-1}(\Pi^n_m)^T$,
	which is the second equality of (\ref{cg.2.6})
\hfill $\Box$

Using a similar argument to continuous time linear system, we have the following result:

\begin{cor}\label{ccg.2.2}
	Consider a continuous time linear system
	\begin{align}\label{cg.2.9}
		\dot{\xi}(t)=A\xi(t),\quad \xi(t)\in \R^n.
	\end{align}
	Its least square projected system on $\R^m$ is
	\begin{align}\label{cg.2.10}
		\dot{x}(t)=A_{\pi}x(t),\quad x(t)\in \R^m,
	\end{align}
	where, ~$A_{\pi}$ is the same as in (\ref{cg.2.6}).
\end{cor}

As an application, assume $n$ is very large, that is, system (\ref{cg.2.3}) is a large scale one. Then we may project it onto a lower dimensional space ${\cal V}_m$, where, ~$m << n$. That is, we have a lower dimensional trajectory to approximate the original one, which might reduce the computational complexity. In the sequel one may see that the projection of lower dimensional system into a higher dimensional vector space is sometimes also necessary.

Similarly, the projection of linear control systems can also be obtained.

\begin{cor}\label{rcg.2.4}
	\begin{itemize}
		\item[(i)] Consider a discrete-time linear control system
		\begin{align}\label{cg.2.11}
			\begin{cases}
				\xi(t+1)=A\xi(t)+Bu,\quad \xi(t)\in \R^n\\
				y(t)=C\xi(t),\quad y(t)\in \R^p.
			\end{cases}
		\end{align}
		Its least square projected system on $\R^m$ is
		\begin{align}\label{cg.2.12}
			\begin{cases}
				x(t+1)=A_{\pi}x(t)+\Pi^n_m B u,\quad x(t)\in \R^m\\
				y(t)=C_{\pi}x(t),
			\end{cases}
		\end{align}
		where, $A_{\pi}$ is defined as in (\ref{cg.2.6}). Moreover,
		\begin{align}\label{cg.2.13}
			C_{\pi}=\begin{cases}
				C(\Pi^n_m)^T\left(\Pi^n_m(\Pi^n_m)^T\right)^{-1},\quad n\geq p\\
				C\left((\Pi^n_m)^T\Pi^n_m\right)^{-1}(\Pi^n_m)^T,\quad n< p.
			\end{cases}
		\end{align}
		
		\item[(ii)] Consider a continuous time linear control system
		\begin{align*}
			\begin{cases}
				\dot{\xi}(t)=A\xi(t)+Bu,\quad \xi(t)\in \R^n\\
				y(t)=C\xi(t),\quad y(t)\in \R^p.
			\end{cases}
		\end{align*}
		Its least square projected system on $\R^m$ is
		\begin{align*}
			\begin{cases}
				\dot{x}(t)=A_{\pi}x(t)+\Pi^n_mBu,\quad x(t)\in \R^m\\
				y(t)=C_{\pi}x(t),\quad y(t)\in \R^p,
			\end{cases}
		\end{align*}
		where $A_{\pi}$ is defined as in (\ref{cg.2.6}),
		$C_{\pi}$ as in (\ref{cg.2.13}).
	\end{itemize}
\end{cor}

\subsection{Approximation of linear dimension-varying systems}

Consider a discrete-time linear dimension-varying system
\begin{align}\label{cg.3.1}
	\xi(t+1)=A(t)\xi(t),
\end{align}
where $\xi(t)\in \R^{n(t)}$, ~$\xi(t+1)\in \R^{n(t+1)}$, ~$A(t)\in {\cal M}_{n(t+1)\times n(t)}$.

We search its least square projection on $\R^m$ as
\begin{align}\label{cg.3.2}
	x(t+1)=A_{\pi}x(t).
\end{align}

Similarly to the constant dimensional case, the following result can be obtained.

\begin{prp}\label{pcg.3.1} Let (\ref{cg.3.2}) be the least square projected system of (\ref{cg.3.1}) on $\R^m$. Then
	\begin{align}\label{cg.3.3}
		A_{\pi}=\begin{cases}
			\Pi^{n(t+1)}_mA(\Pi^{n(t)}_m)^T\left(\Pi^{n(t)}_m (\Pi^{n(t)}_m)^T\right)^{-1}\quad n(t)\geq m\\
			\Pi^{n(t+1)}_mA\left((\Pi^{n(t)}_m)^T\Pi^{n(t)}_m\right)^{-1}(\Pi^{n(t)}_m)^T\quad n(t)< m.
		\end{cases}
	\end{align}
\end{prp}

An obvious advantage of this projection is the projected system is of constant dimension.

Consider a continuous-time linear dimension-varying system

\begin{align}\label{cg.3.301}
	\dot{\xi}(t)=A(t)\xi(t),
\end{align}
where $\xi(t)\in \R^{n(t)}$, ~$\xi(t+1)\in \R^{n(t+1)}$, ~$A(t)\in {\cal M}_{n(k)\times n(k)}$, ~$k\leq t<k+1$.

We search its least square projection on $\R^m$ as
\begin{align}\label{cg.3.302}
	\dot{x}(t)=A_{\pi}x(t).
\end{align}

To use the previous technique, we assume the dimension of $x(t)$ is piecewise constant. Precisely speaking, we assume:
\begin{align}\label{cg.3.303}
\quad \dim(\xi(t))=\dim(\xi(n)),\quad n\leq t< n+1.
\end{align}
Then the following result can be obtained.
\begin{prp}\label{pcg.3.2} Let (\ref{cg.3.302}) be the lease square projected system of (\ref{cg.3.301}) on $\R^m$.
	Under the assumption that (\ref{cg.3.303}) holds, the $A_{\pi}$ is as in (\ref{cg.3.3}).
\end{prp}

Next, we consider dimension-varying linear control systems. Using similar technique, it is easy to obtain the following projected control systems:

\begin{prp}\label{pcg.3.3}
	\begin{itemize}
		\item[(i)] Consider a discrete-time linear dimension-varying control system
		\begin{align}\label{cg.3.6}
			\begin{cases}
				\xi(t+1)=A(t)\xi(t)+B(t)u\\
				y(t)=C(t)\xi(t),
			\end{cases}
		\end{align}
		where ~$\xi(t)\in \R^{n(t)}$, ~$A(t),~B(t)\in {\cal M}_{n(t+1)\times n(t)}$, ~$C(t)\in {\cal M}_{p\times n(t)}$.
		Then its least square projected control system is
		\begin{align}\label{cg.3.7}
			\begin{cases}
				x(t+1)=A_{\pi}(t)x(t)+\Pi^{n(t+1)}_mBu,\quad x(t)\in \R^m\\
				y(t)=C_{\pi}(t)x(t), \quad y(t)\in \R^p,
			\end{cases}
		\end{align}
		where $A_{\pi}$ is defined as in (\ref{cg.3.3}). Moreover,
		\begin{align}\label{cg.3.701}
			C_{\pi}=\begin{cases}
				C(t)(\Pi^{n(t)}_m)^T\left(\Pi^{n(t)}_m(\Pi^{n(t)}_m)^T\right)^{-1},\quad n(t)\geq p\\
				C(t)\left((\Pi^{n(t)}_m)^T\Pi^{n(t)}_m\right)^{-1}(\Pi^{n(t)}_m)^T,\quad n(t)< p.
			\end{cases}
		\end{align}
		
		\item[(ii)] Consider a continuous-time linear dimension-varying control system
		\begin{align}\label{cg.3.8}
			\begin{cases}
				\dot{\xi}(t)=A(t)\xi(t)+B(t)u\\
				y(t)=C(t)\xi(t),
			\end{cases}
		\end{align}
		where $\xi(t)\in \R^{n(t)}$,  $A(t), B(t)\in {\cal M}_{n(k)\times n(k)}$, $k\leq t<k+1$, $C(t)\in {\cal M}_{p\times n(t)}$.
		Assume (\ref{cg.3.303}) holds, then its least square projected system is
		\begin{align}\label{cg.3.9}
			\begin{cases}
				\dot{x}(t)=A_{\pi}(t)x(t)+\Pi^{n(t+1)}_mBu,\quad \quad x(t)\in \R^m\\
				y(t)=C_{\pi}(t)x(t),\quad y(t)\in \R^p,
			\end{cases}
		\end{align}
		where $A_{\pi}$ is defined as in (\ref{cg.3.3}), $C_{\pi}$ as in (\ref{cg.3.701}).
	\end{itemize}
\end{prp}

Later on, it will be seen that the fixed-dimensional projected system is a very useful realization of dimension-varying systems.

In the following an example is presented to depict projected system.

\begin{exa}\label{ecg.3.1}
	Consider a dimension-varying system
	\begin{align}\label{cg.3.10}
		\begin{cases}
			\xi(t+1)=A(t)\xi(t)+B(t)u\\
			y(t)=C(t)\xi(t),
		\end{cases}
	\end{align}
	where
	$$
	\xi(t)\in
	\begin{cases}
		\R^5,\quad t ~\mbox{is even},\\
		\R^4,\quad t ~\mbox{is odd}.
	\end{cases}
	$$
	{\small
	$$
	A(t)=
	\begin{cases}
		A_1=\begin{bmatrix}
			1&0&-1&2&1\\
			2&-2&1&1&-1\\
			1&2&-1&-2&0\\
			0&1&0&-1&2\\
		\end{bmatrix},& t~\mbox{is even},\\
		A_2=\begin{bmatrix}
			0&-1&2&1\\
			2&1&1&-1\\
			1&2&-1&0\\
			0&1&0&-1\\
			1&-1&0&1\\
		\end{bmatrix},& t~\mbox{is odd}.\\
	\end{cases},~
	B(t)=
	\begin{cases}
		B_1=\begin{bmatrix}
			2&1\\
			2&-1\\
			1&2\\
			0&-1\\
		\end{bmatrix},& t~\mbox{is even},\\
		B_2=\begin{bmatrix}
			2&1\\
			1&-1\\
			2&-1\\
			0&-1\\
			1&0\\
		\end{bmatrix},& t~\mbox{is odd}.\\
	\end{cases}
	$$
	$$
	C(t)=
	\begin{cases}
		C_1=\begin{bmatrix}
			-1&2&1&1&-1\\
			2&-1&-2&-1&2\\
		\end{bmatrix},& t~\mbox{is even},\\
		C_2=\begin{bmatrix}
			2&1&2&-1\\
			0&1&0&-2\\
		\end{bmatrix},& t~\mbox{is odd}.\\
	\end{cases}
	$$}
	A straightforward computation shows that
	$$
	\Pi^4_3=(I_3\otimes \J_4^T)(I_4\otimes \J_3)/3
	=\begin{bmatrix}
		1&1/3&0&0\\
		0&2/3&2/3&0\\
		0&0&1/3&1
	\end{bmatrix}.
	$$
	$$
	\Pi^5_3=(I_3\otimes \J_5^T)(I_5\otimes \J_3)/3
	=\begin{bmatrix}
		1&2/3&0&0&0\\
		0&1/3&1&1/3&0\\
		0&0&0&2/3&1
	\end{bmatrix}.
	$$
	Then the projected system becomes
	\begin{align}\label{cg.3.11}
		\begin{cases}
			x(t+1)=A_{\pi}(t)x(t)+B_{\pi}(t)u\\
			y(t)=C_{\pi}(t)x(t),
		\end{cases}
	\end{align}
	where,
	$$
	\begin{array}{llll}
		A(t)=\tilde{A}_1;&B(t)=\tilde{B}_1;& C(t)=\tilde{C}_1& \quad t~\mbox{is even},\\
		A(t)=\tilde{A}_2;&B(t)=\tilde{B}_2;& C(t)=\tilde{C}_2&\quad  t~\mbox{is odd},\\
	\end{array}
	$$
	where
	$$
		\tilde{A}_1=\Pi^4_3 A_1 (\Pi^5_3)^T\left(\Pi^5_3(\Pi^5_3)^T\right)^{-1}=\begin{bmatrix}
			0.9316&   -0.5556&    1.6239\\
			1.4325&   -0.3111&   -0.7214\\
			1.0923&   -0.6000&    0.7077
		\end{bmatrix};
	$$
	$$
		\tilde{A}_2=\Pi^5_3 A_2 (\Pi^4_3)^T\left(\Pi^4_3(\Pi^4_3)^T\right)^{-1}=\begin{bmatrix}
			0.8333&    1.3333&    0.8333\\
			2.0500&    1.2500&   -1.0500\\
			0.9167&   -0.5833&    0.4167
		\end{bmatrix};
	$$
	$$
	\tilde{B}_1=\Pi^4_3 B_1
	=\begin{bmatrix}
		2.6667&    1.3333\\
		2.0000&    2.0000\\
		0.3333&   -0.3333
	\end{bmatrix};~
	\tilde{B}_2=\Pi^5_3 B_2=
	\begin{bmatrix}
		2.6667&    0.3333\\
		2.3333&   -1.6667\\
		1.0000&   -0.6667
	\end{bmatrix};
	$$
	$$
		\tilde{C}_1=C_1(\Pi^5_3)^T\left( \Pi^5_3(\Pi^5_3)^T\right)^{-1}=
		\begin{bmatrix}
			-0.0359&    1.7333&   -0.4974\\
			1.3333&   -2.6667&    1.3333
		\end{bmatrix};
	$$
	$$
		\tilde{C}_2=C_2(\Pi^4_3)^T\left( \Pi^4_3(\Pi^4_3)^T\right)^{-1}=\begin{bmatrix}
			1.7000&    2.0000&   -0.7000\\
			0.0500&    1.2500&   -2.0500
		\end{bmatrix}.
	$$
\end{exa}

\section{Differential Structures on DFMs}\label{s5}

With the notion of continuous functions over DFESs, we proceed to endow differential structures to a DFES, generalizing it to a dimension-free manifold.

\subsection{From DFESs to DFMs}

\begin{dfn}\label{dqs.5.1} Given a fiber bundle $(T,\pi,B)$.
\begin{itemize}
\item[(i)] Let $\emptyset \neq O\subset B$ be an open set of $B$. Then $(\pi^{-1}(O),\pi, O)$
is called the open sub-bundle (over $O$).
\item[(ii)] Let $O_{\lambda}$, $\lambda\in \Lambda$ be an open cover of $B$, that is, $\bigcup_{\lambda\in \Lambda}O_{\lambda}=B$, then $\left\{\left(\pi^{-1}(O_{\lambda}),\pi, O_{\lambda}\right)\right\}_{\lambda\in \Lambda}$
is called an open cover of the fiber bundle $(T,{\pi},B)$.
\end{itemize}
\end{dfn}

\begin{dfn}\label{dqs.5.2} Assume $(T,{\pi},B)$ is a fiber bundle, where both $T$ and $B$ are second countable Hausdorff spaces. $(T,{\pi},B)$ is called a $C^r$ DFEB with $B$ as a $C^r$ DFM, if the following conditions are satisfied.
\begin{itemize}
\item[(i)] There is an open cover $\left\{(W_{\lambda},{\pi},B_{\lambda})\right\}_{\lambda\in \Lambda}$
of $(T,{\pi},B)$.

\item[(ii)] For each $(W_{\lambda},{\pi},B_{\lambda})$ there exists an open sub-bundle $\left({\Pr}^{-1}(O_{\lambda}),{\Pr},O_{\lambda}\right)$ of
    $(\R^{\infty},{\Pr},\Omega)$ with two bijective mappings
$$
\Psi_{\lambda}: W_{\lambda}\ra {\Pr}^{-1}(O_{\lambda}),\quad
\varphi_{\lambda}: B_{\lambda} \ra O_{\lambda},
$$
such that $(W_{\lambda},{\pi},B_{\lambda})$ and $({\Pr}^{-1}(O_{\lambda}),{\Pr},O_{\lambda})$ are fiber bundle isomorphic, that is, diagram (\ref{d4}) commutes.
\begin{align}\label{d4}
	\xymatrix{
		W_{\lambda}\ar[r]^{\Psi_{\lambda}}\ar[d]^{\Pr}&{\Pr}^{-1}(O_{\lambda})\ar[d]^{\pi}\\
		B_{\lambda}	\ar[r]^{\varphi_{\lambda}}&O_{\lambda}
	}
\end{align}

\item[(iii)] Assume $B_{\lambda_1}\bigcap B_{\lambda_2}\neq \emptyset$.
Then $\varphi_2\circ \varphi_1^{-1}: \varphi_1(B_{\lambda_1}\bigcap B_{\lambda_2})\ra \varphi_2(B_{\lambda_1}\bigcap B_{\lambda_2})$
is $C^r$. 
\end{itemize}
\end{dfn}

The following proposition comes from definition immediately.

\begin{prp}\label{pqs.5.301} Let $T\stackrel{\pi}{\longrightarrow} B$ be a $C^r$ DFEB. Set $M_n:=\bigcup_{\lambda\in \Lambda}\Psi_{\lambda}^{-1}(O^n_{\lambda})$,
then $M_n$ is an $n$-dimensional $C^r$ manifold. Moreover, $M=\bigcup_{n=1}^{\infty}M_n$.
\end{prp}

The following  example provides a DFM.

\begin{exa}\label{eqs.5.4} Consider $S_{\infty}:=\bigcup_{n=1}^{\infty} S_n$,
where $S_n$ is the $n$-dimensional unit sphere in $\R^{n+1}$, $n=1,2,\cdots$.
Denote by $P_n=(\underbrace{0,\cdots,0}_n,-1)$ and $Q_n=(\underbrace{0,\cdots,0}_n,1)$ the north and south poles of the $n$-dimensional sphere respectively.

\begin{itemize}
\item[(i)] Set $M_n:=S_n\backslash{P_n}$, and define a mapping $\Psi_n:M_n\ra \R^n$ by $\xi_i=\frac{x_i}{1+x_{n+1}},\quad i=1,2,\cdots,n$.

Define $M=\bigcup_{n=1}^{\infty}M_n$,
and using the inherent topology from $\R^{n+1}$ for $M_n$, and assume $M_n$ are clopen in $M$. Then
the mapping $\Psi: M\ra \R^{\infty}$ is a topological isomorphism.

To see this, we have only to show that $\Psi$ is bijective and $\Psi^{-1}$ is also continuous.
It is clear  from definition that
$$
(\xi_1^2+\cdots+\xi_n^2)(1+x_{n+1})^2=\dsum_{i=1}^nx_i^2.
$$
Then we have
\begin{align}\label{qs.5.5}
\|\xi\|^2(1+x_{n+1})^2+x_{n+1}^2=1.
\end{align}
Solving Equation (\ref{qs.5.5}) and noting that $x_{n+1}\neq -1$ yield
\begin{align}\label{qs.5.6}
x_{n+1}=\frac{1-\|\xi\|^2}{1+\|\xi\|^2},
\end{align}
and
\begin{align}\label{qs.5.7}
x_{i}=(1+x_{n+1})\xi_i,\quad i\in[1,n].
\end{align}

(\ref{qs.5.6})-(\ref{qs.5.7}) show that $\Psi^{-1}$ is also continuous.

Next, for $a,b\in M$ we define
\begin{align}\label{qs.5.701}
a\sim_M b \Leftrightarrow \Psi(a)\lra \Psi (b).
\end{align}
Then we can define a mapping $\psi:M/\sim_M\ra \Omega$ by
\begin{align}\label{qs.5.702}
\Psi (a)=x \Rightarrow \psi(\bar{a}):=\bar{x}.
\end{align}
Because of (\ref{qs.5.701}), (\ref{qs.5.702}) is properly defined.

Finally, we define $\pi_1: M\ra M/\sim_M$ as $\pi_1=\psi^{-1}\circ \Pr\circ \Psi$. Then it is ready to verify that
$(M, \pi_1, M/\sim_M)$ is a DFEB and $M/\sim_S$ is a DFM.

\item[(ii)] Set $N_n:=S_n\backslash{Q_n}$, and define a mapping $\Phi_n:N_n\ra \R^n$ by
\begin{align}\label{qs.5.8}
\eta_i=\frac{x_i}{1-x_{n+1}},\quad i=1,2,\cdots,n.
\end{align}
Similarly to case (i),   from definition (\ref{qs.5.8}) we have
$$
(\eta_1^2+\cdots+\eta_n^2)(1-x_{n+1})^2=\dsum_{i=1}^nx_i^2.
$$
Then we have
\begin{align}\label{qs.5.801}
\|\eta\|^2(1-x_{n+1})^2+x_{n+1}^2=1.
\end{align}
Solving Equation (\ref{qs.5.801}) and noting that $x_{n+1}\neq 1$ yield
\begin{align}\label{qs.5.802}
x_{n+1}=\frac{\|\eta\|^2-1}{1+\|\eta\|^2},
\end{align}
and
\begin{align}\label{qs.5.803}
x_{i}=(1-x_{n+1})\eta_i,\quad i\in[1,n].
\end{align}

(\ref{qs.5.802})(\ref{qs.5.803}) show that $\Phi^{-1}$ is also continuous.

Next, for $x,y\in N$ we define
$$
x\sim_N y \Leftrightarrow \Phi(x)\lra \Phi (y),
$$
and $\pi_2: N\ra N/\sim_N$ as $\pi_2=\phi^{-1}\circ\Pr\circ \Phi$, where $\phi$ can be constructed similarly as for $\psi$. Then
$(N, \pi_2, N/\sim_N)$ is a DFEB and $N/\sim_N$ is a DFM.

\item[(iii)] Consider
$$
S_{\infty}=\bigcup_{n=1}^{\infty}S_n=M\bigcup N.
$$
It is clear that $\{M,N\}$ is an open cover of $S_{\infty}$. Consider $(x_1,\cdots,x_{n+1})\in M\bigcap S_n$. From (\ref{qs.5.6})(\ref{qs.5.7}) we can solve $x_1,\cdots,x_{n+1}$ out as $x_i(\xi)$, $i\in[1,n+1]$. Similarly, for $(x_1,\cdots,x_{n+1})\in N\bigcap S_n$, using (\ref{qs.5.802})(\ref{qs.5.803}) we can also express  $x_1,\cdots,x_{n+1}$ as $x_i(\eta)$, $i=1,\cdots,n+1$.

Now, assume $\bar{\xi}\in M/\sim_M$ and $\bar{\eta}\in N/\sim_N$, $\dim(\bar{\xi})=\dim(\bar{\eta})$, and the smallest elements in $\bar{\xi}$ and $\bar{\eta}$ are $\xi_1$ and $\eta_1$ respectively.
Let $\dim(\xi_1)=\dim(\eta_1)=n$. Then $\bar{\xi}$ is said to be equivalent to $\bar{\eta}$, denoted by $\bar{\xi}\leftrightharpoons \bar{\eta}$, if $x_i(\xi_1)=x_i(\eta_1),\quad \forall i=1,\cdots,n+1$.

It is easy to see that $\leftrightharpoons$ is an equivalence relation. Then we define
\begin{align}\label{qs.5.805}
B:=\left(M/\sim_M \bigcup N/\sim_N\right)/\leftrightharpoons.
\end{align}

Define
$$
\begin{array}{l}
B_1:=\left(M/\sim_M \right)/\leftrightharpoons,\\
B_2:=\left(N/\sim_N \right)/\leftrightharpoons.
\end{array}
$$
Then it is easy to verify that $\{B_1,B_2\}$ is an open cover of $B$.
Moreover, it is ready to see that $\pi_1:M\ra B_1$, $\pi_2:N\ra B_2$ are consistent. Hence $\pi:S_{\infty}\ra B$ can be defined as
\begin{align}\label{qs.5.807}
\pi(x):=
\begin{cases}
\pi_1(x),\quad x\in M,\\
\pi_2(x),\quad x\in N.
\end{cases}
\end{align}
We conclude that $\left(S_{\infty},{\pi}, B\right)$ is a DFEB with $B$ a DFM.
\end{itemize}
\end{exa}

\begin{rem}\label{rqs.5.5} Hereafter, to avoid notational mess, we consider only DFES. In fact, all the following arguments can easily be extended to DFM. Hence in the following the DFES can be considered as a bundle of coordinate chart, which has a set of fixed coordinate frames on each leaves.
\end{rem}

\subsection{Vector fields on DFESs}
First, we define the tangent space of $\Omega$.

\begin{dfn}\label{dqg.1.1} Let $\bar{x}\in \Omega$ and $\dim(\bar{x})=m$. Then the tangent space of $\bar{x}$, called the tangent bundle at $\bar{x}$ and denoted by $T_{\bar{x}}(\Omega)$, is defined by
\begin{align}\label{qg.1.1}
T_{\bar{x}}(\Omega):=\R^{[m,\cdot]}.
\end{align}
\end{dfn}

\begin{rem}\label{rqg.1.2}
\begin{itemize}
\item[(i)] When $\Omega$ is replaced by a DFM, Definition \ref{dqg.1.1} can only be considered as for a given fixed set of coordinate charts.
\item[(ii)] Consider $\Omega$. Then we assume on each leaf of $T_{\bar{x}}(\Omega)$, say $\R^s$, where $s=km$, the coordinate frame is fixed as $(x_1,x_2,\cdots,x_s)$. Then the basis of the leaf is $\{\frac{\pa}{\pa x_1},\frac{\pa}{\pa x_2},\cdots,\frac{\pa}{\pa x_s}\}$. Hence each vector at $T_{\bar{x}}(\Omega)$, denoted be $(a_1,a_2,\cdots,a_s)$, is an operator $\dsum_{i=1}^sa_i\frac{\pa}{\pa x_i}$. Of course, this operator can be extended to coordinate-free form for DFM. But restricting on $\Omega$ can avoid the complexity of expression.
\end{itemize}
\end{rem}

Recall the definition of bundle of coordinate neighborhood of DFES.
(refer to Figure \ref{Figqs.4.1}), it is easily seen that for each $\bar{x}\in \Omega$ the bundle of coordinate neighborhood coincides with its tangent bundle.

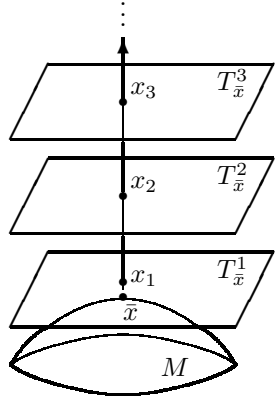
\begin{figure}
\centering
\setlength{\unitlength}{5mm}
\begin{picture}(9,13)\thicklines
\put(1,4){\line(1,0){6}}
\put(2,6){\line(1,0){6}}
\put(1,4){\line(1,2){1}}
\put(7,4){\line(1,2){1}}
\put(1,6.5){\line(1,0){6}}
\put(2,8.5){\line(1,0){6}}
\put(1,6.5){\line(1,2){1}}
\put(7,6.5){\line(1,2){1}}
\put(1,9){\line(1,0){6}}
\put(2,11){\line(1,0){6}}
\put(1,9){\line(1,2){1}}
\put(7,9){\line(1,2){1}}
\put(4,5.3){\line(0,1){1.1}}
\put(4,7.6){\line(0,1){1.3}}
\put(4,10.1){\vector(0,1){1.6}}
\put(3.9,12){$\vdots$}
\qbezier(1,3)(1.5,2.6)(3,2.3)
\qbezier(5,2.3)(6.5,2.6)(7,3)
\qbezier(3,2.3)(4,2.1)(5,2.3)
\qbezier(1,3)(1.5,3.8)(2,4)
\qbezier(7,3)(6.5,3.8)(6,4)
\put(4,4.8){\circle*{0.2}}
\put(4,5.2){\circle*{0.2}}
\put(4,7.5){\circle*{0.2}}
\put(4,10){\circle*{0.2}}
\put(4,4.2){$\bar{x}$}
\put(4.2,5.2){$x_1$}
\put(6.5,5.3){$T_{\bar{x}}^1$}
\put(4.2,7.7){$x_2$}
\put(6.5,7.8){$T_{\bar{x}}^2$}
\put(4.2,10.2){$x_3$}
\put(6.5,10.3){$T_{\bar{x}}^3$}
\put(5,2.7){$M$}
\thinlines
\qbezier(1,3)(1.5,3.4)(3,3.7)
\qbezier(5,3.7)(6.5,3.4)(7,3)
\qbezier(3,3.7)(4,3.9)(5,3.7)
\qbezier(2,4)(4,5.5)(6,4)
\put(4,5){\line(0,1){6.5}}
%

\end{picture}
\caption{Tangent Bundle on Dimension-Free Manifold\label{Figqg.1.1}}
\end{figure}

When a DFM $M$ is considered, Let ~$\bar{x}\in M$ and ~$\dim(\bar{x})=m$, then the tangent bundle $T_{\bar{x}}(M)$ is depicted at Figure \ref{Figqg.1.1}, where $T_{\bar{x}}^i=\R^{im},~i=1,2,\cdots$.
That is, $T_{\bar{x}}(M)=\R^{[m,.]}$.

\vskip 5mm

If we consider the tangent space over whole $\Omega$, that is,
$$
T(\Omega):=\bigcup_{\bar{x}\in \Omega}T_{\bar{x}},
$$
Then it is obvious that $T(\Omega)=\R^{\infty}$.

Next, we define vector fields on $\Omega$. The following definition is also available for DFMs.

\begin{dfn}\label{dqg.1.3} $\bar{X}$ is called a $C^r$ vector field on $\Omega$, denoted by $\bar{X}\in V^r(\Omega)$, if it satisfies the following condition:
\begin{itemize}
\item[(i)] At each point $\bar{x}\in \Omega$, there exists $p=p_{\bar{x}}=\mu_{\bar{x}} \dim(\bar{x})$, called the dimension of the vector field $\bar{X}$ at ~$\bar{x}$ and denoted by $\dim(\bar{X}_{\bar{x}})$, such that $\bar{X}$ assigns to the bundle of coordinate neighborhood  at $\bar{x}$ a $p$ sub-lattice, ~${\cal V}_{O}^{[p,\cdot]}=\{O^{p},O^{2p},\cdots\}$,
then at each leaf of this sub-lattice the vector field assigns a vector $X^{j}\in T_{x_{j \mu}}(O^{jp})$, ~$j=1,2,\cdots$.
\item[(ii)]  ~$\{X^{j}\;|\; j=1,2,\cdots\}$ satisfy consistence condition, that is, $X^j=X^1\otimes \J_j,\quad j=1,2,\cdots$.
\item[(iii)] At each leaf ~$O^{jp}\subset \R^{j\mu \dim(\bar{x})}$,
\begin{align}\label{qg.1.4}
\bar{X}|_{O^{jp}}\in V^r(O^{jp}).
\end{align}
\end{itemize}
\end{dfn}

\begin{dfn}\label{dqg.1.4} A vector field $\bar{X}\in V^r(\Omega)$ is said to be dimension bounded, if $\underset{{\bar{x}\in \Omega}}{\max}\dim(\bar{X}_{\bar{x}})<\infty$.
\end{dfn}

In the following a method is presented to construct a $C^r$ vector field on $\Omega$. The method is similar to the construction of continuous functions. It is first built on ${\cal V}^m=\R^m$, and then extended to  ~$T(\Omega)=\R^{\infty}$.

\begin{alg}\label{aqg.1.5}

\begin{itemize}
\item Step 1: Assume there exists a smallest dimension $m>0$, such that $\bar{X}$ is defined over whole $\R^m$. That is,
\begin{align}\label{qg.1.6}
\bar{X}|_{\R^m}:=X\in V^r(\R^m).
\end{align}
From the constructing point of view: A vector field $X\in V^r(\R^m)$ is firstly given, such that the value of $\bar{X}$ at leaf $\R^m$ is uniquely determined by (\ref{qg.1.6}).

\item Step 2: Extend $X$ to $T_{\bar{y}}$. Assume $\dim(\bar{y})=s$,  denote $m\vee s=t$, $t/m=\a$, $t/s=\b$. Then ~$\dim(T_{\bar{y}})=t$. Let $y\in \bar{y}\bigcap R^{[t,\cdot]}$, and $\dim(y)=kt$, $k=1,2,\cdots$. Define
\begin{align}\label{qg.1.7}
\bar{X}(y):=\Pi^m_{kt}X(\Pi^{kt}_m y),\quad k=1,2,\cdots.
\end{align}
\end{itemize}
\end{alg}

\begin{thm}\label{tqg.1.6}
\begin{itemize}
\item[(i)] The $\bar{X}$ generated by Algorithm \ref{aqg.1.5} is a $C^r$ vector field, that is, $\bar{X}\in V^{r}(\Omega)$.
\item[(ii)] If $\bar{X}\in V^{r}(\Omega)$ is dimension bounded, then $\bar{X}$ can be generated by Algorithm \ref{aqg.1.5}.
\end{itemize}
\end{thm}
\noindent{\it Proof}. 
\begin{itemize}
\item[(i)] By definition, for any $\bar{y}\in \Omega$ and assume $\dim(\bar{y})=s$, then on a sub-lattice $\R^{[t,\cdot]}$  of the bundle of coordinate neighborhood of ~$\bar{y}$ (Since only the DFES is considered now, each leaf of the bundle of coordinate neighborhood can be whole Euclidean space.) a vector $\bar{X}_y$ is assigned. In the following we prove that the set of such vectors are consistent. Assume $\dim(y)=kt=k\b m$, when ~$k=1$,
    $y=y_{\b}$, then
    $$
    \begin{array}{ccl}
    \bar{X}(y_{\b})&=&\Pi^m_{\b m}X(\Pi^{\b m}_m y_{\b})\\
    ~&=&(I_{\b m}\otimes \J_1^T)(I_m\otimes \J_\b)X(\Pi^{\b m}_m y_{\b})\\
    ~&=&(I_m\otimes \J_\b)X(\Pi^{\b m}_m y_{\b})\\
    ~&=& X(\Pi^{\b m}_m y_{\b})\otimes I_{\b}.
    \end{array}
    $$
Similar calculation shows that
    $$
    \bar{X}(y_{k\b})=X(\Pi^{k\b m}_m y_{k\b})\otimes I_{k\b}.
    $$
Since  $y_{\b}\lra y_{k\b}$, then $\Pi^{k\b m}_m y_{k\b}=\Pi^{\b m}_m y_{\b}$.
Hence,
    $$
    \bar{X}(y_{k\b})=\bar{X}(y_{\b})\otimes \J_k.
    $$
The consistence is proved.

    Finally, we show (\ref{qg.1.4}) holds. That is, to show that on leaf $\R^{jp}$, $\bar{X}$ is a $C^r$ vector field. Since on a leaf all the points are of the same dimension, then the construction  (\ref{qg.1.7}) ensures
$\bar{X}|_{\R^{jp}}$ is a $C^r$ vector field.

\item[(ii)] Assume $\bar{X}$ is dimension bounded, set
$$
m:=\lcm\left\{\dim(\bar{X}_{\bar{x}})\;|\;\bar{x}\in \Omega\right\}.
$$
Then it is clear that $X:=\bar{X}|_{\R^m}\in C^r(\R^m)$. Moreover, since $\bar{X}$  satisfies Definition  ~\ref{dqg.1.3}, then starting from this $X$, the vector field constructed by (\ref{qg.1.7}) coincides with $\bar{X}$.
\end{itemize}
\hfill $\Box$

Hereafter, we consider only dimension bounded vector fields. This is because note only they are easily constructible, but also they are practically useful in modeling dynamic systems.

We construct an example.

\begin{exa}\label{eqg.1.7} Let ~$X=(x_1+x_2,x_2^2)^T\in C^{\omega}(\R^2)$. Assume $\bar{X}\in C^{\omega}(\Omega)$ is generated by $X$.
\begin{itemize}
\item[(i)] Consider $\bar{y}\in \Omega$, $\dim(\bar{y})=3$, Denote $y_1=(\xi_1,\xi_2,\xi_3)^T\in \R^3$. Since $2\wedge 3=6$, ~$\bar{X}$ at
    $$
    \bar{y}\bigcap \R^{6k}=\{y_2,y_4,y_6,\cdots\}
    $$
    is well defined.

Now consider  $y_2$.
    $$
    \bar{X}(y_2)=\Pi^2_6X(\Pi^6_2(y_2))=(I_2\otimes \J_3)X\left(\frac{1}{3}(I_2\otimes \J_3^T)(y_1\otimes \J_2)\right)=\begin{bmatrix}
     \frac{2}{3}(\xi_1+\xi_2+\xi_3)\\
     \frac{2}{3}(\xi_1+\xi_2+\xi_3)\\
     \frac{2}{3}(\xi_1+\xi_2+\xi_3)\\
     \frac{1}{9}(\xi_2+2\xi_3)^2\\
     \frac{1}{9}(\xi_2+2\xi_3)^2\\
     \frac{1}{9}(\xi_2+2\xi_3)^2\\
     \end{bmatrix}
     $$

     Consider $y_4$, similar calculation shows that
    $$
    \bar{X}(y_4)=\Pi^2_{12}X(\Pi^{12}_2(y_4))=\bar{X}(y_2)\otimes \J_2.
    $$
In fact, we have
     $$
     \bar{X}(y_{2k})=\bar{X}(y_2)\otimes \J_k,\quad k=1,2,\cdots.
     $$
     \item[(ii)] Consider $\bar{X}|_{\R^6}$:

     Assume  $z=(z_1,z_2,z_3,z_4,z_5,z_6)^T\in \R^6$. Then
     \begin{align}\label{qg.1.8}
     X^6:=\bar{X}_z=\Pi^2_6 X(\Pi^6_2 z)=\begin{bmatrix}
     \frac{1}{3}(z_1+z_2+z_3+z_4+z_5+z_6)\\
     \frac{1}{3}(z_1+z_2+z_3+z_4+z_5+z_6)\\
     \frac{1}{3}(z_1+z_2+z_3+z_4+z_5+z_6)\\
     \frac{1}{9}(z_4+z_5+z_6)^2\\
     \frac{1}{9}(z_4+z_5+z_6)^2\\
     \frac{1}{9}(z_4+z_5+z_6)^2\\
     \end{bmatrix}.
     \end{align}
     ~$X^6\in V^{\omega}(\R^6)$ is a standard vector field.
     \end{itemize}
     \end{exa}

Next, we consider the integral curve of a vector field on $\Omega$.

\begin{dfn}\label{dqg.1.8} Assume ~$\bar{X}\in C^r(\Omega)$,
$X\in C^r(R^n)$ is its generator, if $X=\bar{X}|_{\R^n}$.
The generator of smallest dimension is called the minimum generator.
\end{dfn}

The following result is an immediate consequence of the definition and Theorem \ref{tqg.1.6}.

\begin{prp}\label{pqg.1.9} Assume $\bar{X}\in V^r(\Omega)$.
\begin{itemize}
\item[(i)] If $X\in V^r(\R^n)$ is its generator, then ~$X\otimes \J_s\in V^r(\R^{sn})$ is also its generator.
\item[(ii)] If $X\in V^r(\R^n)$ is its generator, $Y\in V^r(\R^m)$, ~$m<n$ is also its generator,  then $m|n$, and $X=Y\otimes \J_{n/m}$.
\item[(iii)] Assume $\bar{X}\in V^r(\Omega)$ is dimension bounded, then it has at least one generator, and hence has a minimum generator.
\end{itemize}
\end{prp}

\begin{dfn}\label{dqg.1.10} Let ~$\bar{X}\in C^r(\Omega)$. $\bar{x}(t,\bar{x}_0)$ is called the integral curve of $\bar{X}$ with initial value $\bar{x}_0$, denoted by $\bar{x}(t,\bar{x}_0)=\Phi^{\bar{X}}_t(\bar{x}_0)$, if for each initial value $x_0\in \bar{x}_0\bigcap \R^n$, and each generator of $\bar{X}$, denoted by  $X={\bar{X}}|_{\R^n}$, the following condition holds:
\begin{align}\label{qg.1.10}
\Phi^{\bar{X}}_t(\bar{x_0})|_{\R^n}=\Phi^X_t(x_0),\quad t\geq 0.
\end{align}
\end{dfn}

Next, we consider the existence and the properties of integral curve. First, assume $X=\bar{X}|_{\R^n}$ is the smallest generator of $\bar{X}$. Then, all the generators of $\bar{X}$ are $X_k=\bar{X}|_{\R^{kn}}$, ~$k=1,2,\cdots$. Now assume $\bar{x}_0\in \Omega$, $\dim(\bar{x}_0)=j$, and $j\vee n=s$, then,
$$
\bar{x}\bigcap \R^{\ell}\neq \emptyset,
$$
if and only if, $\ell=ks$, $k=1,2,\cdots$. Denote ~$x_0^s=\bar{x}_0\bigcap \R^s$, then
$$
\Phi^{X_s}_t(x_0^s)=\Phi^{\bar{X}}_t(\bar{x}_0)|_{\R^s}.
$$
Moreover,
$$
\Phi^{\bar{X}}_t(\bar{x}_0)|_{\R^{ks}}=\Phi^{X_{ks}}_t(x_0^{ks})=\Phi^{X_{s}}_t(x_0^{s})\otimes\J_k.
$$
Hence, the integral curve of $\bar{X}$ with initial value $\bar{x}_0$ is a set of integral curves defined on the sublattice bundle ~$\R^{[s,\cdot]}=\{\R^{ks}\;|\;k=1,2,\cdots\}$, and they are all equivalent. That is, for any $0\leq k,~k'<\infty$
$$
\Phi^{X_{ks}}_t(x_0^{ks})\lra  \Phi^{X_{k's}}_t(x_0^{k's}),\quad \forall t\geq 0.
$$

\begin{exa}\label{eqg.1.11} Recall Example \ref{eqg.1.7} Let $\bar{X}\in \Omega$ be generated by $X=(x_1+x_2,x_2^2)^T\in C^{\omega}(\R^2)$, and assume the initial value is $\bar{x}_0\in \Omega$, $\dim(\bar{x}_0)=3$, i.e., $x_1=(\xi_1,\xi_2,\xi_3)^T$. Find the integral curve of $\bar{X}$ initiated at $\bar{x}_0$.

Since $2\vee 3=6$, the integral curve is a set of equivalent curves defined on $\R^{6k}$, $k=1,2,\cdots$. We can first calculate the one defined on $\R^6$, ${\cal X}|_{\R^6}:=X^6$, it is calculated by (\ref{qg.1.8}). Note that $x^0_2:=\bar{x}_0\bigcap \R^6$, then $x^0_2=(\xi_1,\xi_1,\xi_2,\xi_2,\xi_3,\xi_3)^T$. Hence the integral curve is $\Phi^{X^6}_t(x^0_2)$.
It follows that
\begin{align}\label{qg.1.11}
\Phi^{\bar{X}}_t(\bar{x}_0)=\left\{\Phi^{X^6}_t(x^0_2)\otimes \J_k\;|\;k=1,2,\cdots\right\}.
\end{align}
\end{exa}

\subsection{Distributions on DFESs}

\begin{dfn}\label{dqg.6.1} A distribution $\bar{D}$ on $\Omega$ is a rule, which assigns at each point $\bar{x}\in \Omega$ a sub-lattice of its bundle of coordinate neighborhood $O_{\bar{x}}$, denoted by  $O_j=O_{\bar{x}}\cap \R^{jrs}$, ~$r\in \Z_+$, ~$s=\dim(\bar{x})$, ~$j=1,2,\cdots$, and on the tangent space of $x_{jr}\in O_{jr}$, $T_{x_{jr}}(\R^{jrs})$, a subspace $D_j(x_{rj})\subset T_{x_{rj}}(\R^{rjs})$. Moreover, this set of subspaces  satisfies the consistence condition, i.e.,
\begin{align}\label{qg.6.1}
D_j(x_{rj})=D_1(x_r)\otimes \J_j,\quad j=1,2,\cdots.
\end{align}
\end{dfn}

Similarly to vector fields, a distribution can be construct as follows: First, a distribution can be defined on a leaf ${\cal V}^m=\R^m$, then it is extended to $T(\Omega)=\R^{\infty}$.

\begin{alg}\label{aqg.6.2}

\begin{itemize}
\item Step 1: Assume $m$ is the smallest one, such that $\bar{D}$ is defined on leaf $\R^m$. That is,
\begin{align}\label{qg.6.2}
\bar{D}|_{\R^m}:=D(x)\subset T^{r}(\R^m).
\end{align}

\item Step 2: Extend $D(x)$ to $T_{\bar{y}}(\Omega)$.  Let $\dim(\bar{y})=s$,  and $m\vee s=t$, $t/m=\a$, $t/s=\b$. Then $\dim(T_{\bar{y}})=t$. Assume $y\in \bar{y}\bigcap R^{[t,\cdot]}$, and $\dim(y)=kt$, $k=1,2,\cdots$. Define
\begin{align}\label{qg.6.3}
D(y):=\Pi^m_{kt}D(\Pi^{kt}_m y),\quad k=1,2,\cdots.
\end{align}
\end{itemize}
\end{alg}

Similarly to vector fields, the following result can be obtained.

\begin{thm}\label{tqg.6.3}
The $\bar{D}$ constructed by Algorithm \ref{aqg.6.2} is a distribution on $\Omega$. That is,
 $\bar{D}(\bar{x})\subset T_{\bar{x}}(\Omega)$, ~$\forall \bar{x}\in \Omega$.
\end{thm}

The most commonly used distributions are expanded by a set of vector fields.

\begin{dfn}\label{dqg.6.4} Assume $\bar{X}_i\in V^r(\Omega)$ and ~$\dim(\bar{X}_i)=m_i$, $i\in[1,n]$, and ~$m=\lcm\{m_i\;|\;i\in [1,n]\}$. Moreover, let $\bar{X}_i|_{\R^m}=X_i$, and $D_m(x)\subset T(\R^m)$ be the distributions generated by the expansions of~$X_i$, $i\in [1,n]$. Then the distribution $\bar{D}\subset T(\Omega)$ constructed by $D_m(x)s$, $s=1,2,\cdots$ is called the distribution spanned by $\bar{X}_i$, ~$i\in[1,m]$.
\end{dfn}

\begin{dfn}\label{dqg.6.4} Assume $\bar{X}_i\in V^{\infty}(\Omega)$, $\dim(\bar{X}_i)=m_i$, $i=1,2$, and $m=m_1\vee m_2$. Then the Lie bracket of $\bar{X}_1$ and $\bar{X}_2$ is defined by
\begin{align}\label{qg.6.4}
[\bar{X}_1,\bar{X}_2]:=\bar{X}\in V^{\infty}(\Omega)
\end{align}
where $\bar{X}$ is the vector field determined by generator $X$, and
$$
X=[\bar{X}_1\big|_{\R^m},\bar{X}_2\big|_{\R^m}].
$$
\end{dfn}

\begin{exa}\label{eqg.6.5} Assume $\bar{X},\bar{Y}\in V^{\infty}(\Omega)$, $\bar{X}$ and $\bar{Y}$ are generated by $X_0\in V^{\infty}(\R^2)$ and $Y_0\in V^{\infty}(\R^3)$, where
\begin{align}\label{qg.6.5}
\begin{array}{l}
X_0(x)=[x_1+x_2,x_2^2]^T,\\
Y_0(y)=[y_1^2,0,y_2+y_3]^T.
\end{array}
\end{align}
Then $m=2\v3=6$. On leaf $\R^6$, we have
$$
X(z):=\Pi^2_6X_0(\Pi^6_2 z)=[\a,\a,\a,\b,\b,\b]^T,
$$
where,
$$
\a=\frac{1}{3}(z_1+z_2+z_3+z_4+z_5+z_6),\quad \b=\frac{1}{9}(z_4+z_5+z_6).
$$
$$
Y(z):=\Pi^3_6Y_0(\Pi^6_3 z)=[\gamma,\gamma,0,0,\mu,\mu]^T,
$$
where,
$$
\gamma=\frac{1}{4}(z_1+z_2)^2,\quad \mu=\frac{1}{2}(z_3+z_4+z_5+z_6).
$$
Then $\bar{Z}:=[\bar{X},\bar{Y}]$, which is generated by $Z_0\in V^{\infty}(\R^6)$, and
$$
\begin{array}{ccl}
Z_0&=&[X,Y]=\frac{\pa Y}{\pa z}X-\frac{\pa X}{\pa z}Y\\
~&=&[a,a,b,c,d,d]^T.
\end{array}
$$
where
$$
\begin{array}{ccl}
a&=&\frac{1}{3}(z_1+z_2)(z_1+z_2+z_3+z_4+z_5+z_6)-\frac{1}{6}(z_1+z_2)^2-\frac{1}{3}(z_3+z_4+z_5+z_6)\\
b&=&-\frac{1}{6}(z_1+z_2)^2-\frac{1}{3}
(z_3+z_4+z_5+z_6)\\
c&=&-\frac{2}{9}(z_4+z_5+z_6)(z_3+z_4+z_5+z_6)\\
d&=&\frac{1}{6}(z_1+z_2+z_3+z_4+z_5+z_6)+\frac{1}{6}(z_4+z_5+z_6)^2\\
~&~&-\frac{2}{9}(z_4+z_5+z_6)(z_3+z_4+z_5+z_6)\\
\end{array}
$$
\end{exa}

\begin{dfn}\label{dqg.6.6}
\begin{itemize}
\item[(i)] Distribution $\bar{D}(\bar{x})\subset T_{\bar{x}}(\Omega)$, $\bar{x}\in \Omega$ is called an involutive distribution, if any two vector fields $\bar{X}, ~\bar{Y}\in \bar{D}$ satisfy $[\bar{X},\bar{Y}]\in \bar{D}$.
\item[(ii)] Let $\bar{X}_i$, ~$i\in [1,n]$ be given. The involutive distribution generated by $\{\bar{X}_i\;|\;i\in [1,n]\}$, or equivalently, the smallest involutive distribution containing $\{\bar{X}_i\;|\;i\in [1,n]\}$, is called the Lie algebra generated by $\{\bar{X}_i\;|\;i\in [1,n]\}$, denoted by $\left< \bar{X}_i\;|\;i\in [1,n] \right>_{LA}$.
\end{itemize}
\end{dfn}

\section{Dimension-free Riemannian Manifolds}\label{s6}

Parallel to the previous discussions, this section introduce the tensor fields and co-distributions on a DFES, and use them to define the Riemannian and symplectic structure over DFESs.

\subsection{Co-vector fields and co-distributions on DFESs}

First, we define the cotangent space on DFES $\Omega$.

\begin{dfn}\label{dqg.2.1} Let $\bar{x}\in \Omega$ and $\dim(\bar{x})=m$. Then the cotangent space at $\bar{x}$, called the cotangent bundle at $\bar{x}$ and denoted by $T^*_{\bar{x}}(\Omega)$, is defined by
	\begin{align}\label{qg.2.1}
		T^*_{\bar{x}}(\Omega):={\cal V}^{*[m,\cdot]}.
	\end{align}
\end{dfn}

\begin{rem}\label{rqg.2.2} When $\Omega$ is replaced by DFM,  Definition \ref{dqg.2.1} remains available.
\end{rem}

Similarly to tangent bundle, for a given $\bar{x}\in \Omega$, each leaf of its cotangent bundle is an Euclidean space. Moreover, the cotangent bundle at each point is a sub-lattice of $\R^{\infty}$. If $\Omega$ is replaced by a DFM, then the cotangent bundle is similar to tangent bundle. Hence, the Figure \ref{Figqg.1.1} may also be considered as a description of cotangent bundle of DFM.

Next, we define co-vector field on $\Omega$. The following definition is also applicable to DFM.

\begin{dfn}\label{dqg.2.3} $\bar{\omega}$ is called a $C^r$ co-vector field on ~$\Omega$, denoted by $\bar{\omega}\in V^{*r}(\Omega)$, if it satisfies the following conditions:
	\begin{itemize}
		\item[(i)] At each point $\bar{x}\in \Omega$, there exists a $p=p_{\bar{x}}=\mu_{\bar{x}} \dim(\bar{x})$, called the dimension of co-vector field $\bar{\omega}$ at $\bar{x}$, denoted by $\dim(\bar{\omega}_{\bar{x}})$, such that $\bar{\omega}$ assigns a $p$-upper sub-lattice of the bundle of coordinate neighborhood at $\bar{x}$ as ~${\cal V}_{O}^{[p,\cdot]}=\{O^{p},O^{2p},\cdots\}$,
		and then a set of co-vectors $\omega^{j}\in T^*_{x_{j \mu}}(O^{jp})$, ~$j=1,2,\cdots$.
		\item[(ii)]  ~$\{\omega^{j}\;|\; j=1,2,\cdots\}$ satisfy consistent condition, that is,
		\begin{align}\label{qg.2.3}
			\omega^j=\omega^1\otimes \frac{1}{j}\J^T_j,\quad j=1,2,\cdots.
		\end{align}
		\item[(iii)] On each leaf $O^{jp}\subset \R^{j\mu \dim(\bar{x})}$,
		\begin{align}\label{qg.2.4}
			\bar{\omega}|_{O^{jp}}\in V^{*r}(O^{jp}).
		\end{align}
	\end{itemize}
\end{dfn}

\begin{dfn}\label{dqg.2.4} ~$\bar{\omega}\in V^{*r}(\Omega)$ is said to be dimension bounded, if
	\begin{align}\label{qg.2.5}
		\max_{\bar{x}\in \Omega}\dim(\bar{\omega}_{\bar{x}})<\infty.
	\end{align}
\end{dfn}

Similarly to the vector field, the co-vector field can be constructed as follows: First, define it on a leaf ${\cal V}^m=\R^m$, then extend it to $T^*(\Omega)=\R^{\infty}$.

\begin{alg}\label{aqg.2.5}
	
	\begin{itemize}
		\item Step 1: Assume there exists a smallest $m$, such that $\bar{\omega}$ is defined on $\R^m$. That is
		\begin{align}\label{qg.2.6}
			\bar{\omega}|_{\R^m}:=\omega\in V^{*r}(\R^m).
		\end{align}
		From constructing point of view, assume $\omega\in V^{*r}(\R^m)$, then $\bar{\omega}$ is defined on $\R^m$ as in (\ref{qg.2.6}).
		
		\item Step 2:  Extend $\omega$ to $T^*_{\bar{y}}$. Assume $\dim(\bar{y})=s$, $m\vee s=t$, $t/m=\a$, $t/s=\b$, and then $\dim(T_{\bar{y}})=t$. Let ~$y\in \bar{y}\bigcap R^{[t,\cdot]}$, and ~$\dim(y)=kt$, ~$k=1,2,\cdots$. Define
		\begin{align}\label{qg.2.7}
			\bar{\omega}(y):=\omega(\Pi^{kt}_m y)\Pi^{kt}_m,\quad k=1,2,\cdots.
		\end{align}
	\end{itemize}
\end{alg}

Similarly to the case of vector fields, we can prove the following theorem:

\begin{thm}\label{tqg.2.6}
	\begin{itemize}
		\item[(i)] The $\bar{\omega}$ constructed by Algorithm \ref{aqg.2.5} is a $C^r$ co-vector field, that is, ~$\bar{\omega}\in V^{*r}(\Omega)$.
		\item[(ii)] If $\bar{\omega}\in V^{*r}(\Omega)$ is dimension bounded, then any $\bar{\omega}$ can be constructed through Algorithm \ref{aqg.2.5}.
	\end{itemize}
\end{thm}

Similarly to vector fields, hereafter we consider only dimension bounded co-vector fields.

Similarly to vector fields, co-vector fields are defined on a sub-lattice of the bundle of coordinate neighborhood of a point $\bar{x}\in \Omega$. Assume $\omega=\bar{\omega}|_{\R^n}$ is the smallest generator of $\bar{\omega}$. Then, all the generators of  $\bar{\omega}$ are $\omega_k=\bar{\omega}|_{\R^{kn}}$, ~$k=1,2,\cdots$. Now assume $\bar{x}_0\in \Omega$ and ~$\dim(\bar{x}_0)=j$. Denote $j\vee n=s$, then,
$$
\bar{x}\bigcap \R^{\ell}\neq \emptyset,
$$
if and only if, $\ell=ks$, $k=1,2,\cdots$.

In fact, a co-vector field can also be considered as a function of vector field. Hence, the consistence of co-vector fields and vector fields is important. The following proposition shows this consistence.

\begin{prp}\label{pqg.2.7} Let ~$\bar{X}\in V^r(\Omega)$, ~$\bar{\omega}\in V^{*r}(\Omega)$, and $\dim(\bar{X})=\dim(\bar{\omega})$. Then at any point $\bar{x}\in \Omega$ and the sub-lattice of the bundle of coordinate neighborhood of $\bar{x}$ where both $\bar{X}$ and $\bar{\omega}$ are well defined, the action of $\bar{\omega}$ on $\bar{X}$, denoted by $\bar{\omega}(\bar{X})$, is uniquely defined. That is, on $x_k=\bar{x}\bigcap \R^{kp}$, ~$k=1,2,\cdots$,
	\begin{align}\label{qg.2.8}
		\bar{\omega}(\bar{X})|_{x_k}=const., \quad k=1,2,\cdots.
	\end{align}
\end{prp}
\noindent{\it Proof}.  Denote $\dim{x}=s$, $\dim(\bar{X})=\dim(\bar{\omega})=m$. According to the previous argument, it is clear that the sub-lattice, where both $\bar{X}$ and $\bar{\omega}$ are defined, is
	$\{x_p,x_{2p},\cdots\}$, where, ~$p=s\vee m$.
	To prove (\ref{qg.2.8}), it is enough to show that
	\begin{align}\label{qg.2.9}
		\bar{\omega}(\bar{X})|_{x_k}=
		\bar{\omega}(\bar{X})|_{x_1}, \quad k=1,2,\cdots.
	\end{align}
	Assume $p=rm$, then
	\begin{align*}
		\begin{array}{l}
			\omega_1=\omega\left(\Pi^{rm}_m(x_p)\right)\Pi^{rm}_m,~
			\omega_k=\omega\left(\Pi^{rkm}_m(x_{kp})\right)\Pi^{rkm}_m,\\
			X_1=\Pi^m_{rm}X\left(\Pi^{rm}_m(x_p)\right),~
			X_k=\Pi^m_{rkm}X\left(\Pi^{rkm}_m(x_{kp})\right),
		\end{array}
	\end{align*}
	Using which we have
	\begin{align}\label{qg.2.11}
		\begin{array}{l}
			\omega_1(X_1)=\omega\left(\Pi^{rm}_m(x_p)\right)\Pi^{rm}_m
			\Pi^m_{rm}X\left(\Pi^{rm}_m(x_p)\right)\\
			\omega_k(X_k)=\omega\left(\Pi^{rkm}_m(x_{kp})\right)\Pi^{rkm}_m
			\Pi^m_{rkm}X\left(\Pi^{rkm}_m(x_{kp})\right)\\
		\end{array}
	\end{align}
	Since $x_p\lra x_{kp}$, then  $\Pi^{rm}_m(x_p)=\Pi^{rkm}_m(x_{kp})$.
	Hence, to prove (\ref{qg.2.11}) it is enough to show
	$$
	\Pi^{rm}_m\Pi^m_{rm}=\Pi^{rkm}_m\Pi^m_{rkm}.
	$$
	A straightforward computation shows
	$$
	\Pi^{rm}_m\Pi^m_{rm}=\Pi^{rkm}_m\Pi^m_{rkm}=1.
	$$
\hfill $\Box$

Co-vector field is also called one-form. Assume $\bar{f}\in C^r(\Omega)$, then on each leaf $R^m$ $f^m:=\bar{f}|_{\R^m}$ has its differential
\begin{align}\label{qg.2.12}
	df^m=(\frac{\pa f^m}{\pa x_1},\frac{\pa f^m}{\pa x_2},\cdots, \frac{\pa f^m}{\pa x_m}).
\end{align}

Then one sees easily that

\begin{prp}\label{pqg.2.8}
	(\ref{qg.2.12}) generates a co-vector field.
\end{prp}
\noindent{\it Proof}. 
	Taking $df^m$ as the smallest generator of a co-vector field. Consider the differential of $\bar{f}$ on $\R^{km}$. Assume $y\in \R^{km}$, consider $f(\Pi^{km}_my)$. A simple computation shows that
	\begin{align}\label{qg.2.13}
			d f(\Pi^{km}_my)=\left(\frac{\pa f^m}{\pa x_1}|_{\Pi^{km}_my},\cdots,
			\frac{\pa f^m}{\pa x_m}|_{\Pi^{km}_my}\right)\Pi^{km}_m=d f(\Pi^{km}_my)\Pi^{km}_my.
	\end{align}
	This fact shows that the differential of $\bar{f}$ on leaf $\R^{km}$ is exactly the co-vector field deduced by ~$df^m$.
\hfill $\Box$

\begin{dfn}\label{dqg.7.1}  A co-distribution $\bar{D}^*$ on $\Omega$ is a rule, which assigns at each point  $\bar{x}\in \Omega$ a sub-lattice $O_j=O_{\bar{x}}\cap \R^{jrs}$, ~$r\in \Z_+$, ~$s=\dim(\bar{x})$, ~$j=1,2,\cdots$, of its bundle of coordinate neighborhood  $O_{\bar{x}}$, and a sub-space $D^*_j(x_{rj})\subset T^*_{x_{rj}}(\R^{rjs})$ at $x_{jr}\in O_{jr}$. Moreover, this set of sub-spaces of $T^*_{x_{jr}}(\R^{jrs})$ satisfy the consistence condition, i.e.,
	\begin{align}\label{qg.7.1}
		D^*_j(x_{rj})=D^*_1(x_r)\otimes \J^T_j,\quad j=1,2,\cdots.
	\end{align}
\end{dfn}

Similarly to distribution, a $C^r$ co-distribution on $\Omega$ can be constructed as follows:

\begin{alg}\label{aqg.7.2}
	
	\begin{itemize}
		\item Step 1: Assume $m$ is the smallest one, such that $\bar{D}^*$ is defined on $\R^m$. That is,
		\begin{align*}
			\bar{D}^*|_{\R^m}:=D^*(x)\subset T^{*r}(\R^m).
		\end{align*}
		
		\item Step 2: Extend  $D^*(x)$ to $T^*_{\bar{y}}(\Omega)$. Assume $\dim(\bar{y})=s$, $m\vee s=t$, $t/m=\a$, $t/s=\b$, and $\dim(T^*_{\bar{y}})=t$. Let $y\in \bar{y}\bigcap R^{[t,\cdot]}$, and $\dim(y)=kt$, $k=1,2,\cdots$. Define
		\begin{align*}
			D^*(y):=D^*(\Pi^{kt}_m y)\Pi^{kt}_m,\quad k=1,2,\cdots.
		\end{align*}
	\end{itemize}
\end{alg}

Similarly to distributions, the most important co-distributions are generated by a set of co-vector fields.

\subsection{Tensor fields on quotient spaces}

 Let
 $$
 \phi:\underbrace{V(\R^m)\times\cdot\times V(\R^m)}_r\times \underbrace{V^*(\R^m)\times\cdot\times V^*(\R^m)}_s\ra \R
 $$ be an $(r,s)$ th order tensor field on $\R^m$, where $r$ is the covariant order and $s$ is the contro-variant order. The set of $(r,s)$ th order tensor fields is denoted by $T^r_s(\R^m)$. Let ~$\{e_1,e_2,\cdots,e_m\}$  be a basis of $V(\R^m)$, and $\{d_1,d_2,\cdots,d_m\}$ be a basis of $V^*(\R^m)$. Then
\begin{align}\label{qg.3.1}
\begin{array}{ccl}
\gamma^{i_1,i_2,\cdots,i_r}_{j_1,j_2,\cdots,j_s}&:=&\phi\left(e_{i_1},e_{i_2}, \cdots,e_{i_r},d_{j_1},d_{j_2},\cdots,d_{j_s}\right),\\
~&~& 1\leq i_1,\cdots,i_r\leq m,\;
1\leq j_1,\cdots,j_s\leq m,
\end{array}
\end{align}
are called the structure parameters of $\phi$. Using structure parameters, the structure matrix is constructed as follows:
\begin{tiny}
\begin{align}\label{qg.3.2}
\Gamma_{\phi}:=
\begin{bmatrix}
\gamma^{11\cdots 1}_{11\cdots 1}&\cdots&\gamma^{11\cdots m}_{11\cdots 1}&\cdots&
\gamma^{mm\cdots 1}_{11\cdots 1}&\cdots&\gamma^{mm\cdots m}_{11\cdots 1}\\
\vdots&\cdots&\vdots&\cdots&
\vdots&\cdots&\vdots\\
\gamma^{11\cdots 1}_{11\cdots m}&\cdots&\gamma^{11\cdots m}_{11\cdots m}&\cdots&
\gamma^{mm\cdots 1}_{11\cdots m}&\cdots&\gamma^{mm\cdots m}_{11\cdots m}\\
\vdots&\cdots&\vdots&\cdots&
\vdots&\cdots&\vdots\\
\gamma^{11\cdots 1}_{mm\cdots 1}&\cdots&\gamma^{11\cdots m}_{mm\cdots 1}&\cdots&
\gamma^{mm\cdots 1}_{mm\cdots 1}&\cdots&\gamma^{mm\cdots m}_{mm\cdots 1}\\
\vdots&\cdots&\vdots&\cdots&
\vdots&\cdots&\vdots\\
\gamma^{11\cdots 1}_{mm\cdots m}&\cdots&\gamma^{11\cdots m}_{mm\cdots m}&\cdots&
\gamma^{mm\cdots 1}_{mm\cdots m}&\cdots&\gamma^{mm\cdots m}_{mm\cdots m}\\
\end{bmatrix}
\end{align}
\end{tiny}

Using this structure matrix,
we have the evaluation formula for $\phi$ as
\begin{align}\label{qg.3.3}
\phi(X_1,\cdots,X_r,\omega_1,\cdots,\omega_s)=
\omega_s \cdots \omega_1\Gamma_{\phi}X_1\cdots X_r.
\end{align}

In the following we construct a tensor field on $\Omega$, denoted by $\bar{\Xi}\in T^r_s(\Omega)$. Assume $\Xi\in T^r_s(\R^m)$ is the smallest generator of $\bar{\Xi}\in T^r_s(\Omega)$, and denote
$$
\bar{\Xi}|_{\R^{km}}:=\Xi_k.
$$
Then it is enough to construct the structure matrix $\Xi_k$, $k=1,2,\cdots$.

It is clear that $\Xi_k$ should satisfy the following requirement: for any $X_1,\cdots,X_r\in V^r(\R^m)$ and $\omega_1,\cdots,\omega_s\in V^{*r}(\R^m)$, their tensor value of $\Xi$ should be the same as the value of $\Xi_k$ with its arguments as the projected vectors and co-vectors to $\R^{km}$. That is,
\begin{align}\label{qg.3.4}
\begin{array}{rl}
~&\Xi(x)(X_1(x),\cdots,X_r(x),\omega_1(x),\cdots,\omega_s(x))\\
=&\Xi_k(y)(\pi^k_{km}(X_1(x(y)),\cdots,\pi^k_{km}(X_r(x(y)),\pi^k_{km}(\omega_1(x(y)),\cdots,\pi^k_{km}(\omega_s(x(y))),
\end{array}
\end{align}
where $y=\Pi^m_{km}(x)$, $x(y)=\Pi^{km}_m(y)$.

Let $\Gamma(x)$ be the structure matrix of $\Xi$ and ~$\Gamma_k(y)$ be the structure matrix of$\Xi_k$.
Then (\ref{qg.3.4}) can be expressed as
$$
\begin{array}{rl}
~&\omega_s(x)\cdots \omega_1(s)\Gamma(x) X_1(x) \cdots X_r(x)\\
=&\omega_s(\Pi^{km}_m y)\Pi^{km}_m \cdots \omega_1(\Pi^{km}_m y)\Pi^{km}_m \Gamma_k(y)\Pi^m_{km}X_1(\Pi^{km}_m(y))\cdots
\Pi^m_{km}X_r(\Pi^{km}_m(y))\\
=&\omega_s(x)\cdots \omega_1(x)\left(I_{(s-1)m}\otimes \Pi^{km}_m\right)\cdots\\
~&\left(I_m\otimes \Pi^{km}_m\right)\Pi^{km}_m \Gamma_k(y) \Pi^m_{km}\left(I_m\otimes \Pi^m_{km}\right)\cdots\\
~&\left(I_{(r-1)m}\otimes \Pi^m_{km}\right)X_1(x)\cdots X_r(x).
\end{array}
$$
Hence we have
\begin{align}\label{qg.3.5}
\begin{array}{ccl}
\Gamma(x)&=&
\left(I_{(s-1)m}\otimes \Pi^{km}_m\right)\cdots\\
~&~&\left(I_m\otimes \Pi^{km}_m\right)\Pi^{km}_m \Gamma_k(y) P^m_{km}\left(I_m\otimes \Pi^m_{km}\right)\cdots\\
~&~&\left(I_{(r-1)m}\otimes \Pi^m_{km}\right).
\end{array}
\end{align}
It follows immediately that
\begin{align}\label{qg.3.6}
\begin{array}{l}
\Gamma_k(y):=
\Pi^{m}_{km}\left(I_m\otimes \Pi^{m}_{km} \right)\cdots\\
~~\left(I_{(s-1)m}\otimes \Pi^{m}_{km} \right)\Gamma(\Pi^{km}_m(y))
\left(I_{(r-1)m}\otimes \Pi^{km}_m\right)\\
~~\cdots \left(I_m\otimes \Pi^{km}_m\right) \Pi^{km}_m.
\end{array}
\end{align}

A straightforward verification shows that the $\Gamma_k$ defined by (\ref{qg.3.6}) satisfies (\ref{qg.3.5}).

Next, set $\bar{x}\in \Omega$, $\dim(\bar{x})=s$, $s\vee m=p$, and let $p=\mu s=\lambda m$. Then, $\bar{\Xi}$ is defined at $\bar{x}\bigcap \R^{kp}$, $k=1,2,\cdots$.  Denote $x_k=\bar{x}\bigcap \R^{kp}$, then
\begin{align}\label{qg.3.7}
\begin{array}{ccl}
\bar{\Gamma}(x_k)&:=&
\Pi^{m}_{k\lambda m}\left(I_m\otimes \Pi^{m}_{k\lambda m} \right)\cdots\\
~&~&\left(I_{(s-1)m}\otimes \Pi^{m}_{k\lambda m} \right)\Gamma(\Pi^{k\lambda m}_m(x_k))\\
~&~&\left(I_{(r-1)m}\otimes \Pi^{k\lambda m}_m\right)\cdots
 \left(I_m\otimes \Pi^{k\lambda m}_m\right) \Pi^{k\lambda m}_m.
\end{array}
\end{align}

\begin{exa}\label{eqg.3.1} Assume $\bar{\Xi}\in T^2_1(\Omega)$, and its smallest generator is $\Xi\in T^2_1(\R^2)$, which has its structure matrix as
\begin{align}\label{qg.3.8}
\Gamma(x)=\begin{bmatrix}
0&\sin(x_1+x_2)&0&\cos(x_1+x_2)\\
-\cos(x_1+x_2)&0,\sim(x_1+x_2)&0\\
\end{bmatrix}
\end{align}
\begin{itemize}

\item[(i)] Find the structure matrix of $\bar{\Xi}|_{\R^4}$.

Using formula (\ref{qg.3.6}), we have
$$
\begin{array}{rl}
\bar{\Xi}|_{\R^4}=&\Pi^2_4\Gamma(\Pi^4_2(y))\left(I_2\otimes \Pi^4_2\right)\Pi^4_2\\
~~=&\Pi^2_4\Gamma([\frac{y_1+y_2}{2},\frac{y_3+y_4}{2}])\left(I_2\otimes \Pi^4_2\right)\Pi^4_2\\
~~=&\frac{1}{4}
\left[\begin{array}{cccccccc}
{\bf 0}&S&{\bf 0}&S&{\bf 0}&C&{\bf 0}&C\\
-C&{\bf 0}&-C&{\bf 0}&S&{\bf 0}&S&{\bf 0}\\
\end{array}
\right],
\end{array}
$$
where%
$$
{\bf 0}=\left\{
\begin{array}{cc}
0&0\\
0&0
\end{array}\right\}
$$
$$
S=\left\{\begin{array}{cc}
\sin\left(\frac{y_1+y_2+y_3+y_4}{2}\right)&\sin\left(\frac{y_1+y_2+y_3+y_4}{2}\right)\\
\sin\left(\frac{y_1+y_2+y_3+y_4}{2}\right)&\sin\left(\frac{y_1+y_2+y_3+y_4}{2}\right)
\end{array}\right\}
$$
$$
C=\left\{\begin{array}{cc}
\cos\left(\frac{y_1+y_2+y_3+y_4}{2}\right)&\cos\left(\frac{y_1+y_2+y_3+y_4}{2}\right)\\
\cos\left(\frac{y_1+y_2+y_3+y_4}{2}\right)&\cos\left(\frac{y_1+y_2+y_3+y_4}{2}\right)
\end{array}\right\}
$$
\item[(ii)]
Assume $\bar{x}\in \Omega$ and $\dim(\bar{z})=3$. Then $\bar{\Xi}$ is defined only on
$z_{kp}$, $p=2\vee 3=6$, $k=1,2,\cdots$. Using formula (\ref{qg.3.7}), we have
$$
\begin{array}{l}
\bar{\Xi}|_{z_2}=\Pi^2_6\Gamma(\Pi^6_2(z_2))\left(I_2\otimes \Pi^6_2\right)\Pi^6_2\\
~~=\Pi^2_6\Gamma([\frac{2z_1+z_2}{3},\frac{z_+2z_3}{3}])\left(I_2\otimes \Pi^6_2\right)\Pi^6_2\\
~~=\frac{1}{9}[A,A,A,B,B,B]\otimes \J_{3\times 3},
\end{array}
$$
where
$$
A=\left\{\begin{array}{cc}
0&\sin\left(\frac{2(z_1+z_2+z_3)}{3}\right)\\
\cos\left(\frac{2(z_1+z_2+z_3)}{3}\right)&0
\end{array}\right\}
$$
$$
B=\left\{\begin{array}{cc}
0&\cos\left(\frac{2(z_1+z_2+z_3)}{3}\right)\\
\sin\left(\frac{2(z_1+z_2+z_3)}{3}\right)&0\\
\end{array}\right\}
$$
\end{itemize}
\end{exa}

\begin{rem}\label{rqg.3.2}
\begin{itemize}
\item[(i)] The above constructing technique is applicable to tensor fields on DFMs. Hence, we assume  tensor fields on DFMs are also properly defined.
\item[(ii)] A tensor field $T^0_s$ is called a covariant tensor field. A tensor field $T_0^r$ is also called an $r$-form.
\end{itemize}
\end{rem}

\subsection{Dimension-free Riemannian manifolds and dimension-free symplectic manifolds}

\begin{dfn}\label{dqg.4.1} Let $\Xi\in T^r_0(\Omega)$ be an $r$ th order covariant tensor field.
\begin{itemize}
\item[(i)] ~$\Xi$ is said to be symmetric, if
\begin{align}\label{qg.4.1}
\Xi(X_1,\cdots,X_r)=\Xi(X_{\sigma(1)},\cdots,X_{\sigma(r)}),\quad \sigma\in {\bf S}_r.
\end{align}
\item[(ii)] ~$\Xi$ is said to be skew-symmetric, if
\begin{align}\label{qg.4.1}
\Xi(X_1,\cdots,X_r)=\sign(\sigma)\Xi(X_{\sigma(1)},\cdots,X_{\sigma(r)}),\quad \sigma\in {\bf S}_r.
\end{align}
\end{itemize}
\end{dfn}

Second order covariant tensor fields (or $2$-forms) are of special importance. Their structure matrices can also be expressed into a quadratic form as
\begin{align}\label{qg.4.2}
M_{\Xi}=\begin{bmatrix}
\gamma^{11}&\gamma^{12}&\cdots&\gamma^{1n}\\
\gamma^{21}&\gamma^{22}&\cdots&\gamma^{2n}\\
\vdots&~&~&~\\
\gamma^{n1}&\gamma^{n2}&\cdots&\gamma^{nn}\\
\end{bmatrix}
\end{align}

Using this structure matrix, the tensor field can be the expressed into a classical quadratic form as
\begin{align}\label{qg.4.3}
\Xi(X_1,X_2)=X_1^TM_{\Xi}X_2.
\end{align}
Then we have

\begin{prp}\label{pqg.4.2} ~$\bar{\Xi}\in T^2(\Omega)$  is symmetric (skew-symmetric), if and only if,
its smallest generator $\Xi$ is symmetric (skew-symmetric). That is, it has a symmetric (skew-symmetric) structure matrix.
\end{prp}

\begin{dfn}\label{dqg.4.3} Consider $\Omega$. Assume there is an order 2 covariant tensor field $\bar{\Xi}\in T^2(\Omega)$.
\begin{itemize}
\item[(i)] $(\Omega,\bar{\Xi})$ is called a dimension-free Riemannian manifold, if
$\bar{\Xi}$ is generated by $\Xi_n=\bar{\Xi}|_{\R^n}$ and
 ~$(\R^n, \Xi_n)$ is a Riemannian manifold.  That is, ~$\Xi_n:=\bar{\Xi}|_{\R^n}\in T^2(\R^n)$ has a symmetric positive definite structure matrix $M_{\Xi_n}$.
\item[(ii)] ~$(\Omega,\bar{\Xi})$ is called a dimension-free symplectic manifold, if $\bar{\Xi}$ is generated by $\Xi_n=\bar{\Xi}|_{\R^n}$, where  ~$n=2m$ is even, and $(\R^n, \Xi_n)$ is a symplectic manifold.  That is, $\Xi_{2m}:=\bar{\Xi}|_{\R^{2m}}\in T^2(\R^{2m})$ has a skew-symmetric, non-singular structure matrix $M_{\Xi_{2m}}$, and $\Xi$ is closed.
\end{itemize}
\end{dfn}

\begin{rem}\label{rqg.4.4} An $r$ form $\Xi$ is closed if $d\Xi=0$ \cite{boo86}. Let $\Xi\in T^2(\R^n)$. $\Xi$ is closed, if and only if, its structure coefficients satisfy\cite{guo05}
\begin{align}\label{qg.4.4}
\frac{\pa}{\pa x_i}(\gamma^{jk})+\frac{\pa}{\pa x_j}(\gamma^{ki})+\frac{\pa}{\pa x_k}(\gamma^{ij})=0,\quad 1\leq ~i,~j,~k\leq n.
\end{align}
\end{rem}

\begin{dfn}\label{dqg.4.401}
\begin{itemize}
\item[(i)] A DFM $M$ with a two-form $\bar{\Theta}$ is called a dimension-free Riemannian manifold, if there exists an open sub-bundle cover of $M$ such that each open sub-bundle is bundle isomorphic to an open sub-bundle of a Riemannian manifold of $\Omega$ with Riemannian two-form $\bar{\Xi}$.
Moreover, $\bar{\Theta}$ is isomorphic consistently to $\bar{\Xi}$.
\item[(ii)] A DFM $M$ with a two-form $\bar{\Theta}$ is called a dimension-free symplectic manifold, if there exists an open sub-bundle cover of $M$ such that each open sub-bundle is bundle isomorphic to an open sub-bundle of a symplectic manifold of $\Omega$ with symplectic two-form $\bar{\Xi}$.
Moreover, $\bar{\Theta}$ is isomorphic consistently to $\bar{\Xi}$.
\end{itemize}
\end{dfn}

\begin{rem}\label{rqg.4.402} Let $M_1$ and $M_2$ be two DFMs, and $\Psi:M_1\ra M_2$ be an isomorphism.
$\bar{\Xi}_i\in T^r_s(M_i)$, $i=1,2$ are said to be isomorphic consistently, if
\begin{align}\label{qg.4.401}
\begin{array}{l}
\bar{\Xi}_1(X_1,\cdots,X_r;\omega_1,\cdots,\omega_s)\\
=
\bar{\Xi}_2(\Psi_*(X_1),\cdots,\Psi_*(X_r),{\Psi^{-1}}^*(\omega_1),\cdots,{\Psi^{-1}}^*(\omega_s)),\\
~~~~~~~~~~X_1,\cdots,X_r\in V(M_1),\; \omega_1,\cdots,\omega_s\in V^*(M_1).
\end{array}
\end{align}
\end{rem}

\begin{exa}\label{eqg.4.5} Consider DEES $\Omega$.
\begin{itemize}
\item[(i)] Assume $\Omega$ has a two form $\bar{\sigma}\in T^2(\Omega)$ with its generator $\sigma\in T^2(\R^2)$, and the structure matrix of $\sigma$ is $M_{\sigma}=\begin{bmatrix}
0&-1\\
1&0
\end{bmatrix}$.
Since $M_{\sigma}$ is a symplectic matrix, it is clear that $(\Omega,\bar{\sigma})$ is a dimension-free symplectic manifold.

 Assume $\bar{x}\in \Omega$ and $\dim(\bar{x})=2$. Then $\bar{\sigma}$ is defined on $T_{\bar{x}}=\{\R^{2k}\;|\;k=1,2,\cdots\}$. Moreover, the structure matrix of $\bar{\sigma}(x_k)$ is
\begin{align*}
M_{k}:=M|_{x_k}=\Pi^{2}_{2k}M_{\sigma}\Pi^{2k}_2=M_{\sigma}\otimes I_k.
\end{align*}

 Assume $\bar{y}\in \Omega$ and $\dim(\bar{y})=3$. Then $\bar{\sigma}$ is defined on $T_{\bar{y}}=\{\R^{6k}\;|\;k=1,2,\cdots\}$. Moreover, the structure matrix of $\bar{\sigma}(y_{2k})$ is
\begin{align*}
M_{k}:=M|_{y_{2k}}=\Pi^{2}_{6k}M_{\sigma}\Pi^{6k}_2=M_{\sigma}\otimes I_{3k}.
\end{align*}

\item[(ii)] Assume $\Omega$ has a two form $\bar{\omega}\in T^2(\Omega)$, where $\omega$ is deduced from part of sphere $S_2\backslash{P_2}$. We refer to Example \ref{eqs.5.4} for notations. Assume $S_2$ has the standard distance inherited from $\R^3$. Then the structure matrix of $\omega$ is
\begin{align*}
M_{\omega}=
(\frac{\pa x}{\pa \xi})^TI_3 (\frac{\pa x}{\pa \xi})=
\begin{bmatrix}
\left\|\frac{\pa x}{\pa \xi_1}\right\|^2&\left<\frac{\pa x}{\pa \xi_1},\frac{\pa x}{\pa \xi_2} \right>\\
\left<\frac{\pa x}{\pa \xi_1},\frac{\pa x}{\pa \xi_2} \right>&\left\|\frac{\pa x}{\pa \xi_2}\right\|^2
\end{bmatrix},
\end{align*}
where
$$
\begin{array}{l}
\left\|\frac{\pa x}{\pa \xi_1}\right\|^2=\left(\frac{\pa x_1}{\pa \xi_1}\right)^2+\left(\frac{\pa x_2}{\pa \xi_1}\right)^2+\left(\frac{\pa x_3}{\pa \xi_1}\right)^2\\
\left<\frac{\pa x_1}{\pa \xi},\frac{\pa x_2}{\pa \xi} \right>=
\frac{\pa x_1}{\pa \xi_1}\frac{\pa x_1}{\pa \xi_2}+\frac{\pa x_2}{\pa \xi_1}\frac{\pa x_2}{\pa \xi_2}+\frac{\pa x_3}{\pa \xi_1}\frac{\pa x_3}{\pa \xi_2}\\
\left\|\frac{\pa x}{\pa \xi_2}\right\|^2=\left(\frac{\pa x_1}{\pa \xi_2}\right)^2+\left(\frac{\pa x_2}{\pa \xi_2}\right)^2+\left(\frac{\pa x_3}{\pa \xi_2}\right)^2,
\end{array}
$$
and
$$
\begin{array}{l}
\frac{\pa x_1}{\pa \xi_1}=\frac{2-2\xi_1^2+2\xi^2_2}{(1+\xi_1^2+\xi_2^2)^2}, \quad
\frac{\pa x_1}{\pa \xi_2}=\frac{-4\xi_1\xi_2}{(1+\xi_1^2+\xi_2^2)^2}\\
\frac{\pa x_2}{\pa \xi_1}=\frac{-4\xi_1\xi_2}{(1+\xi_1^2+\xi_2^2)^2}, \quad
\frac{\pa x_2}{\pa \xi_2}=\frac{2+2\xi_1^2-2\xi^2_2}{(1+\xi_1^2+\xi_2^2)^2}\\
\frac{\pa x_3}{\pa \xi_1}=\frac{-4\xi_1}{(1+\xi_1^2+\xi_2^2)^2},\quad
\frac{\pa x_3}{\pa \xi_2}=\frac{-4\xi_2}{(1+\xi_1^2+\xi_2^2)^2}\\
\end{array}
$$
Hence, $(\Omega,\bar{\omega})$ is a dimension-free Riemannian Manifold (DFRM).

Assume $\Theta$ has a two form $\bar{\theta}\in T^2(\Theta)$, where $\theta$ is deduced from part of sphere $S_2\backslash{Q_2}$. Assume $S_2$ has the standard distance inherited from $\R^3$. Then the structure matrix of $\Theta$ is
\begin{align*}
M_{\theta}=
(\frac{\pa x}{\pa \eta})^TI_3 (\frac{\pa x}{\pa \eta})=
\begin{bmatrix}
\left\|\frac{\pa x}{\pa \eta_1}\right\|^2&\left<\frac{\pa x}{\pa \eta_1},\frac{\pa x}{\pa \eta_2} \right>\\
\left<\frac{\pa x}{\pa \eta_1},\frac{\pa x}{\pa \eta_2} \right>&\left\|\frac{\pa x}{\pa \eta_2}\right\|^2
\end{bmatrix},
\end{align*}
where
$$
\begin{array}{l}
\left\|\frac{\pa x}{\pa \eta_1}\right\|^2=\left(\frac{\pa x_1}{\pa \eta_1}\right)^2+\left(\frac{\pa x_2}{\pa \eta_1}\right)^2+\left(\frac{\pa x_3}{\pa \eta_1}\right)^2\\
\left<\frac{\pa x_1}{\pa \eta},\frac{\pa x_2}{\pa \eta} \right>=
\frac{\pa x_1}{\pa \eta_1}\frac{\pa x_1}{\pa \eta_2}+\frac{\pa x_2}{\pa \eta_1}\frac{\pa x_2}{\pa \eta_2}+\frac{\pa x_3}{\pa \eta_1}\frac{\pa x_3}{\pa \eta_2}\\
\left\|\frac{\pa x}{\pa \eta_2}\right\|^2=\left(\frac{\pa x_1}{\pa \eta_2}\right)^2+\left(\frac{\pa x_2}{\pa \eta_2}\right)^2+\left(\frac{\pa x_3}{\pa \eta_2}\right)^2,
\end{array}
$$
and
$$
\begin{array}{l}
\frac{\pa x_1}{\pa \eta_1}=\frac{2-2\eta_1^2+2\eta^2_2}{(1+\eta_1^2+\eta_2^2)^2}, \quad
\frac{\pa x_1}{\pa \eta_2}=\frac{-4\eta_1\eta_2}{(1+\eta_1^2+\eta_2^2)^2}\\
\frac{\pa x_2}{\pa \eta_1}=\frac{-4\eta_1\eta_2}{(1+\eta_1^2+\eta_2^2)^2}, \quad
\frac{\pa x_2}{\pa \eta_2}=\frac{2+2\eta_1^2-2\eta^2_2}{(1+\eta_1^2+\eta_2^2)^2}\\
\frac{\pa x_3}{\pa \eta_1}=\frac{-4\eta_1}{(1+\eta_1^2+\eta_2^2)^2}, \quad
\frac{\pa x_3}{\pa \eta_2}=\frac{-4\eta_2}{(1+\eta_1^2+\eta_2^2)^2}\\
\end{array}
$$
Hence, $(\Theta,\bar{\theta})$ is also a DFRM.

 Combining $\Omega$ with $\Theta$, one sees that $S_{\infty}$ is a DFRM too.
\end{itemize}
\end{exa}

\section{Dimension-Varying Dynamic (Control) Systems}\label{s7}

\subsection{Dynamic (control) systems over DFMs}

\subsubsection{Projection of dynamic (control) systems}

Consider a dynamic system over $\R^p$, described as
\begin{align}\label{pds.1}
\Sigma:~\dot{x}=F(x),\quad x\in \R^p.
\end{align}

\begin{dfn}\label{dpds.1}
Consider dynamic system (\ref{pds.1}). Its projection onto $\R^q$ is a dynamic system over $\R^q$, described as
\begin{align}\label{pds.2}
\pi^p_q(\Sigma):~\dot{z}=\tilde{F}(z),\quad z\in \R^q,
\end{align}
where
\begin{align}\label{pds.3}
\tilde{F}(z)=\Pi^p_qF(\Pi^q_p(z)).
\end{align}
\end{dfn}

Consider a control system
\begin{align}\label{pds.4}
\Sigma^C:~\dot{x}=F(x,u),\quad x\in \R^p,\; u\in \R^r.
\end{align}

\begin{dfn}\label{dpds.101}
Consider control system (\ref{pds.4}).
The $u=u_1,\cdots,u_r$ can be considered as parameters. Then its projection to $R^q$ can still be considered as a projection of vector field  as
\begin{align}\label{pds.5}
\pi^p_q(\Sigma^C):~\dot{z}=\tilde{F}(z,u),\quad z\in \R^q,\;u\in \R^r,
\end{align}
where
\begin{align}\label{pds.6}
\tilde{F}(z,u)=\Pi^p_qF(\Pi^q_p(z),u).
\end{align}
\end{dfn}

\begin{rem}\label{rpds.2}
The projection from $\R^p$ to $\R^q$ can be extended to a projection from $p$ dimensional manifold to $q$ dimensional manifold.
    Then the above descriptions can be considered as the expression over local coordinate charts.
\end{rem}

The following is an example.

\begin{exa}\label{epds.3} Consider the following control system $\Sigma$:
\begin{align}\label{pds.7}
\begin{cases}
\dot{x}_1=u_1\sin(x_1+x_2),\\
\dot{x}_2=u_2\cos(x_1+x_2).
\end{cases}
\end{align}
\begin{itemize}
\item[(i)] Project (\ref{pds.7}) onto $\R^3$.
It is ready to calculate that
$$
\Pi^3_2=\frac{1}{3}\begin{bmatrix}
2&1&0\\
0&1&2
\end{bmatrix}, \quad
\Pi^2_3=\frac{1}{2}\begin{bmatrix}
2&0\\
1&1\\
1&2
\end{bmatrix}.
$$
Then the projected system $\pi^2_3(\Sigma)$ is calculated as
\begin{align}\label{pds.8}
\begin{cases}
\dot{z}_1=u_1\sin(\frac{2}{3}(z_1+z_2+z_3)),\\
\dot{z}_2=\frac{1}{2}\left(u_1\sin(\frac{2}{3}(z_1+z_2+z_3))+u_2\cos(\frac{2}{3}(z_1+z_2+z_3))\right),\\
\dot{z}_3=u_2\cos(\frac{2}{3}(z_1+z_2+z_3)).
\end{cases}
\end{align}
\item[(ii)] Project (\ref{pds.7}) onto $\R^4$.
We have
$$
\Pi^4_2=\frac{1}{2}\begin{bmatrix}
1&1&0&0\\
0&0&1&1
\end{bmatrix}, \quad
\Pi^2_4=\begin{bmatrix}
1&0\\
1&0\\
0&1\\
0&1
\end{bmatrix}.
$$
Then the projected system $\pi^2_4(\Sigma)$ is easily obtained as
\begin{align}\label{pds.9}
\begin{cases}
\dot{z}_1=u_1\sin(\frac{1}{2}(z_1+z_2+z_3+z_4)),\\
\dot{z}_2=u_1\sin(\frac{1}{2}(z_1+z_2+z_3+z_4)),\\
\dot{z}_3=u_2\cos(\frac{1}{2}(z_1+z_2+z_3+z_4)),\\
\dot{z}_4=u_2\cos(\frac{1}{2}(z_1+z_2+z_3+z_4)).
\end{cases}
\end{align}
\item[(iii)] Project (\ref{pds.8}) (i.e., $\pi^2_3(\Sigma)$) back to $\R^2$, we have
\begin{align}\label{pds.10}
\begin{cases}
\dot{x}_1=\frac{1}{6}\left(5u_1\sin(x_1+x_2)+\cos(x_1+x_2)\right),\\
\dot{x}_2=\frac{1}{6}\left(u_1\sin(x_1+x_2)+5\cos(x_1+x_2)\right).\\
\end{cases}
\end{align}
System (\ref{pds.10}) differs from the original system, which means the transfer loses information.

\item[(iii)] Project (\ref{pds.9}) (i.e., $\pi^2_4(\Sigma)$) back to $\R^2$, we have $\Sigma$, which means the transfer is lossless.
\end{itemize}
\end{exa}

Motivated by the above example, we can prove the following result.

\begin{prp}\label{ppds.4} Let $f(x)\in V^{\infty}(\R^p)$ and $q=kp$. Then
\begin{align}\label{pds.11}
\pi^q_p\circ \pi^p_q(f(x))=f(x).
\end{align}

\end{prp}
\noindent{\it Proof}.  First, a straightforward computation can prove the following equality:
\begin{align}\label{pds.12}
\Pi^{kp}_p \Pi^p_{kp}=I_p.
\end{align}
Using it, we have that
$$
\begin{array}{lcl}
f(x)&\xrightarrow{\pi^p_{kp}}&\Pi^p_{kp}f\left(\Pi^{kp}_p z\right)\\
~&\xrightarrow{\pi^{kp}_{p}}&\Pi^{kp}_p\Pi^p_{kp}
f\left(\Pi^{kp}_p\Pi^p_{kp} x\right)=f(x).
\end{array}
$$

\hfill $\Box$

\begin{rem}\label{rpds.5}
\begin{itemize}
\item[(i)] Proposition \ref{ppds.4} shows that when a vector field is projected onto its multiple-dimension Euclidean space there is no information losing. This is essential for constructing a control system on dimension-free manifolds.
\item[(ii)] In previous sections, according to the definition of a vector field on $\R^{\infty}$,  for a vector field on $\R^p$ only its integral curves over $\R^{kp}$ are considered. That means only the projection of the vector field to $\R^{kp}$ are considered. In current definition, the projection to ant $\R^s$ is allowed. In fact, only when $s=kp$, the extension is lossless. When $s\neq kp$, the projected system can only be considered as an approximated system of the original one. Its integral curve can not be considered as the integral curve of the original system, but only an approximation too.
\end{itemize}
\end{rem}

\subsubsection{Nonlinear control systems over $\Omega$}

To avoid counting the degrees of differentiability, the functions, vector fields, etc. are assumed to be of $C^{\infty}$.

\begin{dfn}\label{dcs.1.1}
\begin{itemize}
\item[(i)] A nonlinear control system over ~$\Omega$, denoted by ~$\bar{\Sigma}$, is described by
\begin{align}\label{cs.1.1}
\begin{cases}
\dot{\bar{x}}=\overline{F}(\bar{x},u)\\
\bar{y}_s=\bar{h}_s(\bar{x}),\quad s\in[1,p],
\end{cases}
\end{align}
where ~$\overline{F}(\bar{x},u)\in V^{\infty}(\Omega)$, $\bar{h}_s\in C^{\infty}(\Omega)$, $s\in[1,p]$ , ~$u=(u_1,u_2,\cdots,u_m)$ are controls, which can be considered as parameters in ~$F$.  $\bar{y}_s$, $s\in[1,m]$ are outputs.
\item[(ii)] Let ~$\bar{f},~\bar{g}_j,~j\in[1,m]\in V^{\infty}(\Omega)$,
\begin{align}\label{cs.1.2}
\begin{cases}
\overline{F}(\bar{x},u)=\bar{f}(\bar{x})+\dsum_{j=1}^m\bar{g}_j(\bar{x})u_j,\\
\bar{y}_s=\bar{h}_s(\bar{x}),\quad s\in[1,p].
\end{cases}
\end{align}
Then ~(\ref{cs.1.2}) is called an affine nonlinear control system over ~$\Omega$.
\item[(iii)] Assume
$$
q:=\lcm\left(\dim(\overline{F}(u)),~\dim(\bar{h}_s),~s\in[1,p]\right).
$$
Then ~$\bar{\Sigma}|_{\R^q}:=\Sigma$ is called the minimum generator of ~$\bar{\Sigma}$, denoted by
\begin{align}\label{cs.1.3}
\begin{cases}
\dot{x}=F(x,u),\quad x\in \R^q\\
y_s=h_s(x),\quad s\in[1,p].
\end{cases}
\end{align}
\item[(iv)] $\bar{\Sigma}$ is said to be completely controllable (observable), if~ $\Sigma$ is completely controllable (observable).
\end{itemize}
\end{dfn}

\begin{rem}\label{rcs.1.2}
\begin{itemize}
\item[(i)] If the state space of minimum generator is on~$\R^q$, then, $\bar{\Sigma}$ is well posed on ~$\R^{kq}$, $k=1,2,\cdots$. They will be called the realizations of $\overline{F}(u)$. Unfortunately, the control properties, such as controllability, observability, etc., of the realizations with different dimensions are not the same. Hence the controllability and observability of $\bar{\Sigma}$ are defined by corresponding properties of its minimum generator.
\item[(ii)] Hereafter all the control properties of $\bar{\Sigma}$ are referred to the corresponding properties of its minimum generator.
\end{itemize}
\end{rem}

\subsection{Linear systems over DFMs}

\begin{itemize}
\item Linear vector fields
\end{itemize}

Let $\bar{X}\in V^{\infty}(\Omega)$ be a linear vector field and ~$\dim(\bar{X})=m$. Then there exists $A\in {\cal M}_{m\times m}$ such that $X:=\bar{X}|_{\R^m}=Ax$. Consider $\bar{X}|_{\R^{km}}$:
Let $y\in \R^{km}$. Then
\begin{align}\label{cs.3.1}
X_k:=\bar{X}(y)=\Pi^{m}_{km}(X\left(\Pi^{km}_m(y)\right))=\Pi^{m}_{km}A\Pi^{km}_my:=A_ky,
\end{align}
where,
\begin{align}\label{cs.3.2}
A_k=\Pi^{m}_{km}A\Pi^{km}_m=\frac{1}{k}\left(I_m\otimes \J_k\right)A\left(I_m\otimes \J^{\mathrm{T}}_k\right).
\end{align}

Then we consider the integral curve of $\bar{X}$.

Assume $\bar{X}\in V^{\infty}(\Omega)$ is a linear vector field and $\dim(\bar{X})=m$. $X:=\bar{X}|_{\R^m}=Ax$. Consider $\bar{x}^0\in \Omega$.
\begin{itemize}
\item[] Case $1$: Assume $\dim(\bar{x}^0)=m$. Then the integral curve of $\bar{X}$ with initial value $\bar{x}^0$ is defined only on a filter of the tangent bundle
$$
T_{\bar{x}^0}\bigcap \R^{km},\quad k=1,2,\cdots.
$$

On~$T_{\bar{x}^0}\bigcap \R^{km}$, at $x^0_k=x^0_1\otimes \J_k$, the vector field is determined by (\ref{cs.3.2}). Then the integral curve becomes
\begin{align*}
\begin{aligned}
x_k(t,x^0_k)&=e^{X_kt}x^0_k\\
&\textstyle=\left(I_{km}+t(I_m\otimes \J_k)A(I_m\otimes \J_k^{\mathrm{T}})+\right.\\
~&\left.\frac{t^2}{2!} (I_m\otimes \J_k)A^2(I_m\otimes \J_k^{\mathrm{T}})+\cdots\right)(x^0_1\otimes \J_k)\\
~&=\frac{1}{k}(I_m\otimes \J_k)e^{At}(I_m\otimes \J_k^{\mathrm{T}})(x^0_1\otimes \J_k)\\
~&=(I_m\otimes \J_k)e^{At}x_0\\
~&=e^{At}x_0\otimes \J_k.
\end{aligned}
\end{align*}

\item[]   Case~$2$: Assume $\dim(\bar{x}^0)=s$, $m\vee s=p=km=rs$. Then the integral curve of $\bar{X}$ with initial value $\bar{x}^0$ is defined on a filter of its tangent bundle
$$
T_{\bar{x}^0)}\bigcap \R^{jp},\quad j=1,2,\cdots.
$$
On leaf ~$T_{\bar{z}^0}\bigcap \R^{p}$, the initial value  ~$z^0_r=z^0_1\otimes \J_r$, where ~$z^0_1\in \bar{z}^0$ is its smallest element. The vector field is $A_kz$, where $A_k$ is determined by (\ref{cs.3.2}).
Hence, the integral curve is
\begin{align*}
z_r(t,z^0_r)=e^{X_kt}z^0_r=\frac{1}{k}(I_m\otimes \J_k)e^{At}(I_m\otimes J_k^{\mathrm{T}})(z^0_1\otimes \J_r).
\end{align*}
On leaf $T_{\bar{z}^0}\bigcap \R^{jp}$, the integral curve with initial value $z^0_{jr}=z^0_1\otimes \J_{jr}$ is
\begin{align*}
z_{jr}(t,z^0_{jr})&=e^{X_{jk}t}z^0_{jr}=\frac{1}{k}(I_m\otimes \J_k)e^{At}(I_m\otimes J_k^{\mathrm{T}})(z^0_1\otimes \J_{jr}).
\end{align*}
\end{itemize}

Summarizing the above argument, we have the following result.

\begin{prp}\label{pcs.3.1} Let ~$\bar{X}\in V^{\infty}(\Omega)$ be a linear vector field, and $\dim(\bar{X})=m$. $X:=\bar{X}|_{\R^m}=Ax$. Assume ~$\bar{x}^0\in \Omega$, $\dim(\bar{x}^0)=s$.
\begin{itemize}
\item[(i)] If ~$s=m$, then the integral curve of $\bar{X}|_{\R^m}$ is
\begin{align*}
\Phi^X_t(x^0_1)=e^{Xt}x^0_1.
\end{align*}
Hence, the integral curve of~$\bar{X}|_{\R^{rm}}$ becomes
\begin{align*}
\Phi^{X_r}_t(x^0_r)=\left[e^{Xt}x^0_1\right]\otimes \J_r.
\end{align*}
Finally the integral curve of $\bar{X}$ with initial value $\bar{x}^0$ is~$\overline{\Phi^X_t(x^0_1)}\subset \Omega$.
\item[(ii)] If ~$s=km$,  then the integral curve of~$\bar{X}|_{\R^{km}}$ is
\begin{align*}
\Phi^{X_k}_t(x^0_1)=e^{X_kt}x^0_1,
\end{align*}
where , $X_k$ is determined by (\ref{cs.3.1}).
Hence the integral curve of ~$\bar{X}$ with initial value ~$\bar{x}^0$ is~$\overline{\Phi^{X_k}_t(x^0_1)}\subset \Omega$.
\item[(iii)] If~$m\vee s=p=km=rs$, then the integral curve of ~$\bar{X}|_{\R^p}$ is
\begin{align*}
\Phi^{X_k}_t(x^0_r)=e^{X_kt}(x^0_1\otimes I_s).
\end{align*}
Hence, the integral curve of~$\bar{X}$ with initial value ~$\bar{x}^0$ is ~$\overline{\Phi^{X_k}_t(x^0_1\otimes I_s)}\subset \Omega$.
\end{itemize}
\end{prp}

\begin{itemize}
\item Linear control systems
\end{itemize}

First, we consider the relationship among equivalent matrices, equivalent vectors, and linear vector fields.
\begin{dfn}\label{dcs.2.2}
Let ~$A,~B\in {\cal M}$.
\begin{itemize}
\item[(i)] The matrices $A$ and $B$ are said to be type 1 left equivalent, denoted by $A\sim  B$,  if there exist $I_{\a}$ and  $I_{\b}$, such that $A\otimes I_{\a}=B\otimes I_{\b}$.
The equivalence class of $A$ is denoted by $\A_{\ell}$.
\item[(iii)] The matrices $A$ and $B$ are said to be type 2 left equivalent, denoted by $A\approx  B$,  if there exist $J_{\a}$ and  $J_{\b}$, such that $A\otimes J_{\a}=B\otimes J_{\b}$, where $J_i:=\frac{1}{i}\J_{i\times i}$, $i=1,\cdots$.
The equivalence class of $A$ is denoted by $\langle\langle A\rangle\rangle_{\ell}$.
\end{itemize}
\end{dfn}
More general notions on matrix equivalence can be found in \cite{che20}. With these concepts we can define the linear vector fields over $\Omega$.

\begin{prp}\label{pcs.4.1} Let ~$\bar{X}\in V^{\infty}(\Omega)$ be a linear vector field and ~$\dim(\bar{X})=m$. $X:=\bar{X}|_{\R^m}=Ax$. Assume ~$\bar{x}^0\in \Omega$, $\dim(\bar{x}^0)=s$. $m\vee s=p=km=rs$. Then  ~$\bar{X}$ is defined only on the filter of its tangent bundle
$$
x^0_{jr}=T_{\bar{x}^0}\bigcap \R^{jp},\quad j=1,2,\cdots.
$$
Moreover, on the leaf containing~$x^0_{r}$ it is $\bar{X}(x^0_r)=A_kx^0_r$, 
where~$A_k$ is determined by (\ref{cs.3.1}).
On the leaf containing $x^0_{jr}$ it is $\bar{X}(x^0_r)=A_{jk}x^0_{jr},~j=1,2,\cdots$,
where the two sets of consistent matrices are
\begin{align}\label{cs.4.3}
A_{jk}=A_k\otimes I_j\sim A_k,
\end{align}
\begin{align}\label{cs.4.4}
A_{jk}=A_k\otimes J_j\approx A_k,
\end{align}
respectively.
The available variables are
\begin{align}\label{cs.4.5}
x^{0}_{jr}=x^0_r\otimes \J_j\lra x^0_r.
\end{align}
\end{prp}
\noindent{\it Proof}.  In fact, what do we need to show is that the tangent vectors on the bundle leaves are consistent. That is,
\begin{align}\label{cs.4.6}
\bar{X}(x^0_{jr})=\bar{X}(x^0_r)\otimes \J_j,\quad j=1,2,\cdots.
\end{align}
It is obvious that (\ref{cs.4.3}) together with (\ref{cs.4.5}), or (\ref{cs.4.4}) together with (\ref{cs.4.5}) can ensure (\ref{cs.4.6}) to be true.
\hfill $\Box$

Next, we consider the linear control system on $\Omega$. Recall a classical linear system \cite{kai80}
\begin{align}\label{cs.4.7}
\begin{cases}
\dot{x}=Ax+\dsum_{i=1}^mb_iu_i,\\
y=Cx,\quad x\in \R^n, \;y\in \R^p.
\end{cases}
\end{align}
One sees that a classical linear control system consists of three ingredients: linear vector field $Ax$, a set of constant vector fields $B=\{b_1,\cdots,b_m\}$, and linear function $Cx$. To extend a classical linear control system to~ $\Omega$, it is enough to create these three kind of objects to ~$\Omega$. The key of this extension is to make them consistent at each ~$\bar{x}\in \Omega$.

\begin{itemize}
\item[(i)] Linear Vector Field:
Assume the smallest generator of linear vector field $\bar{X}$ is $X=Ax\in V^{\infty}(\R^m)$. $\dim(\bar{x}^0)=s$, $m\vee s=p=\mu m=rs$. Then according to the argument in previous subsection, we know
\begin{align}\label{cs.4.8}
\bar{X}\bigcap T_{\bar{x}^0}=\{\bar{X}(x_{jr})\,|\,j=1,2,\cdots\}.
\end{align}
Moreover,
\begin{align}\label{cs.4.9}
\bar{X}(x_{jr})=A_{j\mu}x_{jr},\quad j=1,2,\cdots,
\end{align}
where, $A_{j\mu}$ is defined by (\ref{cs.3.2}) with $k=j\mu$.

\item[(ii)] Constant Vector Field:
Assume the smallest generator of the constant vector field ~$\bar{X}$ is~$X=b\in V^{\infty}(\R^m)$, where $\dim(\bar{x}^0)=s$, $m\vee s=p=\mu m=rs$, that is, (\ref{cs.4.8}) holds, and
$$
\bar{X}(x_{jr})=\Pi^m_{j\mu m}X(\Pi^{j\mu m}_m x_{jr})=\Pi^m_{j\mu m}b=b\otimes \J_{j\mu}.
$$

\item[(iii)] Linear Function: Let $\bar{h}\in C^{\infty}(\Omega)$. $\dim(\bar{x}^0)=m$, $\bar{h}$ is expressed at $x^0_1$ as $hx=c_mx$, where $c_m^{\mathrm{T}}\in R^m$. Let $\bar{z}\in \Omega$, $\dim(\bar{z})=s$, $m\vee s=p=rs=\mu m$. Then $\bar{h}$ is expressed at $z_1$ as
$$
\bar{h}(z_1)=\bar{h}(\Pi^s_mz_1)=\frac{1}{\mu}c_m\left(I_m\otimes \J^{\mathrm{T}}_{\mu}\right)\left(I_s\otimes \J_{r}\right)z_1
$$
Hence $\bar{h}$ can be expressed on leaf $\R^{s}$ as
\begin{align*}
\bar{h}|_{\R^s}=c_sz,
\end{align*}
where $c_s=\frac{1}{\mu}c_m\left(I_m\otimes \J^{\mathrm{T}}_{\mu}\right)\left(I_s\otimes \J_{r}\right)$.

Particularly, when $s=km$, we have $c_{km}=\frac{1}{k}c_m\left(I_m\otimes \J^{\mathrm{T}}_{k}\right)$.
\end{itemize}

\begin{dfn} \label{dcs.4.2} Assume $\bar{f}(x)$ is a linear vector field, $\bar{B}=[\bar{b}_1,\cdots,\bar{b}_m]$ is a set of constant vector fields, $\bar{C}=[\bar{c}_1,\cdots,\bar{c}_p]^{\mathrm{T}}$ is a set of linear functions, then
\begin{align*}
\begin{cases}
\dot{\bar{x}}=\bar{f}(x)+\bar{B}u,\\
\bar{y}=\bar{C}\bar{x},
\end{cases}
\end{align*}
is a linear control system over $\Omega$.
\end{dfn}

\begin{exa}\label{ecs.4.3} Consider a linear control system  $\bar{\Sigma}$ over $\Omega$, which has its dynamic equation as (\ref{cs.1.2}), where the smallest generator of $\bar{f}$ is  $f(x)=2[x_1+x_2,x_2]^{\mathrm{T}}\in V^{\infty}(\R^2)$, $m=2$, the smallest generators of $\bar{g}_1$ and $\bar{g}_2$ are $g_1=[1,0,0,1]^{\mathrm{T}}\in V^{\infty}(\R^4)$,  $g_2=[0,1,0,0]^{\mathrm{T}}\in V^{\infty}(\R^4)$ respectively. $p=1$, $\bar{h}|_{\R^2}=x_2-x_1$.

Then, $q=4$.
$$
\bar{f}|_{\R^4}=\Pi^2_4f\left(\Pi^4_2[z_1,z_2,z_3,z_4]^{\mathrm{T}}\right)=\begin{bmatrix}
z_1+z_2+z_3+z_4\\
z_1+z_2+z_3+z_4\\
z_3+z_4\\
z_3+z_4\\
\end{bmatrix}:=Az,
$$
where,
$$
A=\begin{bmatrix}
1&1&1&1\\
1&1&1&1\\
0&0&1&1\\
0&0&1&1\\
\end{bmatrix}.
$$
$$
\bar{h}|_{\R^4}=h(\Pi^4_2z)=h(z_1+z_2,z_3+z_4)=z_1+z_2-z_3-z_4:=Cz,
$$
where $C=[1,1,-1,-1]$.

Then the smallest generator of system  $\bar{\Sigma}$, denoted by $\Sigma:=\bar{\Sigma}|_{\R^4}$, is
$$
\begin{cases}
\dot{z}=Az+Bu,\\
y=Cz.
\end{cases}
$$
It is easy to calculate that the controllability matrix of  $\Sigma$ is
$$
{\cal C}=\begin{bmatrix}
 1&     0&     2&     1&     6&     2&    16&     4\\
 0&     1&     2&     1&     6&     2&    16&     4\\
 0&     0&     1&     0&     2&     0&     4&     0\\
 1&     0&     1&     0&     2&     0&     4&     0\\
\end{bmatrix}.
$$
Since $\rank({\cal C})=4$, $\Sigma$ is completely controllable. By definition,  ~$\bar{\Sigma}$ is completely controllable.

The observability matrix of $\Sigma$ is
$$
{\cal O}=\begin{bmatrix}
     1&     1&    -1&    -1\\
     2&     2&     0&     0\\
     4&     4&     4&     4\\
     8&     8&    16&    16\\
\end{bmatrix}
$$
Since $\rank({\cal O})=2<4$,  $\Sigma$ is not completely observable, and so is $\bar{\Sigma}$.
\end{exa}

\subsection{Dimension-varying dynamic (control) systems}

Consider a continuous time dynamic system
\begin{align}\label{dv.1.1}
\dot{x}=F(x), \quad x\in {\cal X},
\end{align}
where $F$ is considered as a vector field on a manifold ${\cal X}$.
Then the solution (integral curve) is expressed as $x(t,x_0)=\Phi^{F}_t(x_0)$.
It is well known that if (\ref{dv.1.1}) is a dynamic system, then $x(t,x_0)$ must be continuous with respect to $t$ \cite{pal82}. Hence a continuous time dimension-varying dynamic system can not be defined on ESDD $\R^{\infty}$. It can only be defined on DFES $\Omega$ (or in general, DFM).

\begin{dfn}\label{ddv.1.1} Consider a dynamic system
\begin{align}\label{dv.1.3}
\dot{\bar{x}}=\bar{F}(\bar{x}), \quad \bar{x}\in \Omega.
\end{align}
\begin{align}\label{dv.1.4}
\dot{x}=F(x), \quad x\in \R^n\subset \R^{\infty},
\end{align}
is called a realization (or a lifting) of (\ref{dv.1.3}), if for each $\bar{x}$ there exists $x\in \bar{x}$, such that
the corresponding vector field $F(x)\in \bar{F}(\bar{x})$.
Meanwhile, system (\ref{dv.1.3}) is called the project system of (\ref{dv.1.4}).
\end{dfn}

The following result is an immediate consequence of the definition.

\begin{prp}\label{pdv.1.2}  $\bar{x}(t)=\bar{x}(t,\bar{x}_0)$ is the solution of (\ref{dv.1.3}), if and only if, $x(t)=x(t,x_0)$ is the solution of (\ref{dv.1.4}), where $x(t)\in \bar{x}(t)$, $t\in [0,\infty)$.
\end{prp}

It is obvious that the lifting system of (\ref{dv.1.3}) is not unique. Assume at $t\in [0,T)$ the system (\ref{dv.1.3}) is lifted to $R^n$ and at $t\in [T,\infty)$ the system is lifted to $\R^m$ ($m\neq n$), then the overall lifting system becomes a dimension-varying system.

\begin{dfn}\label{ddv.1.3} System (\ref{dv.1.4}) is called a dimension-varying system, if there are at least two points $x_1,~x_2$ such that $F(x_1)\in V(\R^{d_1})$, $F(x_2)\in V(\R^{d_2})$ and $d_1\neq d_2$.
\end{dfn}

\begin{rem}\label{rdv.1.4} The Definitions \ref{ddv.1.1} and \ref{ddv.1.3} can easily be extended to corresponding control systems in a natural way. The Proposition \ref{pdv.1.2} has also its corresponding version for control systems.
\end{rem}

In the following we consider how to construct dimension-varying control systems.
We consider two cases.

\begin{itemize}
\item  Case 1: switching dimension-varying control systems
\end{itemize}

Assume the original control system is
\begin{align}\label{dv.2.1}
\dot{x}=F(x,u),\quad x\in \R^m,\;u\in \R^p.
\end{align}
The target system is
\begin{align}\label{dv.2.2}
\dot{z}=G(z,v),\quad z\in \R^n,\;v\in \R^q.
\end{align}

Our purpose is to switch system (\ref{dv.2.1}) to system (\ref{dv.2.2}) at time $t=T$.
To get a continuous trajectory over $\Omega$, the following condition is necessary:
\begin{align}\label{dv.2.3}
\bar{x}(T)=\bar{z}(T)\in \Omega.
\end{align}

\begin{prp}\label{pdv.2.1}
Assume (\ref{dv.2.3}) is satisfied, and assume system (\ref{dv.2.1}) is controllable. Then the dynamic switching from system (\ref{dv.2.1}) to system (\ref{dv.2.2}) at time $t=T$ is realizable.
\end{prp}
\noindent{\it Proof}. 
We construct the following system over $\Omega$:
\begin{align}\label{dv.2.4}
\dot{\bar{\xi}}=\begin{cases}
\bar{F}(\bar{\xi},u),\quad t<T,\\
\bar{G}(\bar{\xi},v),\quad t>T.
\end{cases}
\end{align}
Since system (\ref{dv.2.1}) is controllable, there exists $u(t)$, $t<T$, such that (\ref{dv.2.4}) is controllable to $\bar{\xi}(T)=\bar{x}(T)=\bar{z}(T)$, 
where $\dim(\bar{\xi})=p\wedge q$.

Then the minimum realization of (\ref{dv.2.4}) becomes the required dimension-varying system.
\hfill $\Box$

\begin{exa}\label{edv.2.2}
Consider two systems
\begin{align*}
	\begin{array}{rl}
		\Sigma_1:&\quad\dot{x}=\begin{bmatrix}
			0&1\\
			0&0\end{bmatrix}x+\begin{bmatrix}
			0\\1\end{bmatrix},\quad x\in \R^2,~x(0)=(0,0)^T,\\
		\Sigma_2:&\quad\dot{z}=Az+Bv,\quad z\in \R^3.
	\end{array}
\end{align*}
Design a control such that $\Sigma_1$ is switched to $\Sigma_2$ at $T=1$.

Since $\Sigma_1$ is completely controllable, and $2\wedge 3=1$, so we have to design a control which can drive the system from $x(0)$ to $x(T)$ with $\dim(\bar{x}(T))=1$. We may choose $x(T)=(1,1)^T$. Then it is easy to calculate that the controllability Gramian matrix is
$$
W_C(t)=\int_0^t e^{-A\tau}BB^Te^{-A^T\tau}d\tau=\frac{1}{6}\begin{bmatrix}
2t^3&-3t^2\\-3t^2&t
\end{bmatrix}.
$$
Then the control is
$$
u(t)=-B^Te^{-A^Tt}W_C^{-1}(T)\left(x(0)-e^{-A T}x(T)\right)=-6t.
$$
Using this control, the system can be switched from $\Sigma_1$ to $\Sigma_2$ at $T=1$.
\end{exa}

\begin{itemize}
\item  Case 2: continuous dimension-varying control systems
\end{itemize}

In this case we require the designed dimansion-varying system  has continuous $\bar{F}(\bar{x},u)$. For instance,  in a docking/undocking process, we want the dimension-transient process to be as smooth as possible.

First, let us see what a dimension-varying system with ``continuous" vector field means.

Let $\bar{v}_0$ and $\bar{v}_2$ be two vector fields on $\Omega$. Our purpose is to design a new vector field which continuously transfer from $\bar{v}_0$ to $\bar{v}_2$. Define
$$
\bar{v}:=
\begin{cases}
\bar{v}_0,\quad t\in [t_0,0,t_1),\\
\bar{v}_1=(1-\lambda)\bar{v}_0+\lambda\bar{v}_2,\quad t\in (t_1,t_2),\\
\bar{v}_2,\quad t\in (t_2,\infty),
\end{cases}
$$
where $\lambda=\frac{t-t_1}{t_2-t_1}$.

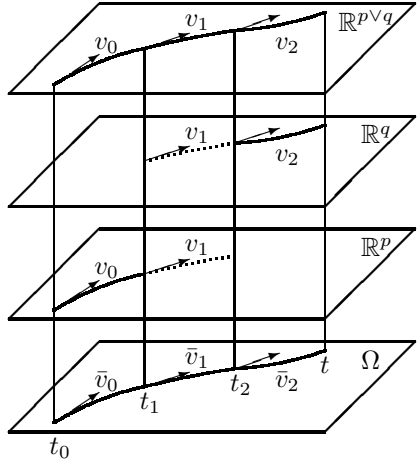
\begin{figure}
\centering
\setlength{\unitlength}{6mm}
\begin{picture}(10,10.5)(0.5,0.5)\thicklines
\put(1,1){\line(1,0){7}}
\put(1,1){\line(1,1){2}}
\put(10,3){\line(-1,-1){2}}
\put(10,3){\line(-1,0){7}}
\put(1,3.5){\line(1,0){7}}
\put(1,3.5){\line(1,1){2}}
\put(10,5.5){\line(-1,-1){2}}
\put(10,5.5){\line(-1,0){7}}
\put(1,6){\line(1,0){7}}
\put(1,6){\line(1,1){2}}
\put(10,8){\line(-1,-1){2}}
\put(10,8){\line(-1,0){7}}
\put(1,8.5){\line(1,0){7}}
\put(1,8.5){\line(1,1){2}}
\put(10,10.5){\line(-1,-1){2}}
\put(10,10.5){\line(-1,0){7}}
\qbezier(2,1.2)(3,1.8)(4,2)
\qbezier(4,2)(5,2.25)(6,2.4)
\qbezier(6,2.4)(7,2.45)(8,2.8)
\qbezier(2,3.7)(3,4.3)(4,4.5)
\qbezier(6,7.4)(7,7.45)(8,7.8)
\qbezier(2,8.7)(3,9.3)(4,9.5)
\qbezier(4,9.5)(5,9.75)(6,9.9)
\qbezier(6,9.9)(7,9.95)(8,10.3)
\put(1.9,0.5){$t_0$}
\put(3.9,1.5){$t_1$}
\put(5.9,1.9){$t_2$}
\put(7.9,2.3){$t$}
\put(8.8,2.4){$\Omega$}
\put(8.8,4.9){$\R^p$}
\put(8.8,7.4){$\R^q$}
\put(8.3,9.9){$\R^{p\vee q}$}
\put(2.9,2){$\bar{v}_0$}
\put(2.9,4.5){$v_0$}
\put(2.9,9.5){$v_0$}
\put(4.9,2.5){$\bar{v}_1$}
\put(4.9,5){$v_1$}
\put(4.9,7.5){$v_1$}
\put(4.9,10){$v_1$}
\put(6.9,2){$\bar{v}_2$}
\put(6.9,7){$v_2$}
\put(6.9,9.5){$v_2$}
\qbezier[15](4,4.5)(5,4.75)(6,4.9)
\qbezier[15](4,7)(5,7.25)(6,7.4)
\thinlines
\put(2,1.2){\line(0,1){7.5}}
\put(4,2){\line(0,1){7.5}}
\put(6,2.4){\line(0,1){7.5}}
\put(8,2.8){\line(0,1){7.5}}
\put(2,1.2){\vector(3,2){1}}
\put(2,3.7){\vector(3,2){1}}
\put(2,8.7){\vector(3,2){1}}
\put(4,2){\vector(3,1){1}}
\put(4,4.5){\vector(3,1){1}}
\put(4,7){\vector(3,1){1}}
\put(4,9.5){\vector(3,1){1}}
\put(6,2.4){\vector(3,1){1}}
\put(6,7.4){\vector(3,1){1}}
\put(6,9.9){\vector(3,1){1}}
\end{picture}
\caption{Lift and projection of integral curves\label{Fig.2.1}}
\end{figure}

Assume the minimum realization of $\bar{v}_0$ is $v_0\in V^r(\R^p)$, the  minimum realization of $\bar{v}_2$ is $\v_2\in V^r(\R^q)$. Then the minimum realization of $\bar{v}_1$ is $v_1\in V^r(\R^{p\vee q})$. Then the integral curve of $\bar{v}$ can be lifted as shown in Fig \ref{Fig.2.1}.

\begin{itemize}
\item Docking:
\end{itemize}

Assume there are two original control systems as
\begin{align}\label{5.6}
\Sigma_1:\quad\dot{x}=F(x,u),\quad x\in \R^m,\;u\in \R^p;
\end{align}
\begin{align}\label{5.7}
\Sigma_2:\quad\dot{z}=G(z,v),\quad z\in \R^n,\;u\in \R^q.
\end{align}
They will be docked into
the target system $\Omega$ as
\begin{align}\label{5.8}
\dot{\xi}=H(\xi,w),\quad \xi\in \R^s,\;w\in \R^{\ell}.
\end{align}

It is required that the docking happens during a transient period $[T_0,T_1]$, and the process is smooth.

\begin{dfn}\label{ddv.3.1} System (\ref{5.6}) and system (\ref{5.7}) are said to be docked into (\ref{5.8}) smoothly during the transient period $[T_0,T_1]$. If the following requirements are satisfied.
\begin{itemize}
\item[(i)] There exists a smooth monotonically non-decreasing function
$\lambda(t)$, $t\in [T_0,T_1]$, such that
$$
\lambda(t)=\begin{cases}
0,\quad t=T_0,\\
1,\quad t=T_1.
\end{cases}
$$
\item[(ii)] There exists a control deformation function $w=\varphi(u,v)$,
and using it a control system over $\Omega$, called a transient system, is constructed as
\begin{align}\label{5.9}
\dot{\bar{\xi}}=(1-\lambda(t))
\begin{bmatrix}
\bar{F}(\bar{x},u)\\
\bar{G}(\bar{z},v)
\end{bmatrix}+\lambda(t)\bar{H}(\bar{\xi},w)+\Psi(\bar{x},\bar{z},\bar{\xi},u,v),\quad t\in [T_0,T_1].
\end{align}
\item[(iii)] The transient system is controllable to
$$
\begin{bmatrix}
\bar{x}(T_1)\\
\bar{z}(T_1)
\end{bmatrix}=\bar{\xi}(T_1),
$$
with $\dim(\bar{\xi(T_1)})=p\wedge q\wedge s$.
\end{itemize}
\end{dfn}

\begin{rem}\label{rdv.3.2}
In (\ref{5.9}) $\Psi(\bar{x},\bar{z},\bar{\xi},u,v)$ is a (fictitious) virtual force, caused by the interaction of docking objects, satisfying
$$
\Psi(\bar{x},\bar{z},\bar{\xi},u,v)=0,\quad if
\begin{bmatrix}
\bar{x}\\
\bar{z}
\end{bmatrix}\lvminus \bar{\xi}=0.
$$
In control of Clutch System the virtual force is chosen as \cite{ser04,tem18}
$$
\tau_c=FcR_a\psi(\omega_i,\omega_0).
$$
Particularly, it was chosen as \cite{che20a}: $\tau_c=\sign(\omega_i-\omega_0)F$.
\end{rem}

\begin{itemize}
\item Undocking:
\end{itemize}

Assume there is an original system  $\Sigma_1$ as:
\begin{align}\label{5.10}
\dot{\xi}=H(\xi,w),\quad \xi\in \R^s,\;w\in \R^{\ell}.
\end{align}
It will be undocked into two systems as
\begin{align}\label{5.11}
\Sigma_2:\quad\dot{x}=F(x,u),\quad x\in \R^m,\;u\in \R^p;
\end{align}
\begin{align}\label{5.12}
\Sigma_3:\quad\dot{z}=G(z,v),\quad z\in \R^n,\;u\in \R^q.
\end{align}
It is required that the docking happens during a transient period $[T_0,T_1]$, and the process is smooth.

\begin{dfn}\label{ddv.3.3}  System (\ref{5.10}) is said to be un-docked into (\ref{5.11}) and (\ref{5.12}) smoothly during the transient period $[T_0,T_1]$. If the follow requirements are satisfied.
\begin{itemize}
\item[(i)] There exists a smooth monotonically non-decreasing function
$\lambda(t)$, $t\in [T_0,T_1]$, such that
$$
\lambda(t)=\begin{cases}
0,\quad t=T_0,\\
1,\quad t=T_1.
\end{cases}
$$
\item[(ii)] There exist two control deformation functions $u=\varphi(w),~ v=\varphi (w)$,
and using them a control system over $\Omega$, called a transient system, is constructed as
\begin{align}\label{5.13}
\begin{array}{rcl}
\dot{\bar{\xi}}&=&(1-\lambda(t))\bar{H}(\bar{\xi},w)\\
~&~&
+\lambda(t)\begin{bmatrix}
\bar{F}(\bar{x},u)\\
\bar{G}(\bar(z),v)
\end{bmatrix}+\Psi(\bar{x},\bar{z},\bar{\xi},w),~ t\in [T_0,T_1].
\end{array}
\end{align}

\item[(iii)] The transient system is controllable to $\bar{\xi}(T_1)=\begin{bmatrix}
\bar{x}(T_1)\\
\bar{z}(T_1)
\end{bmatrix}$,
with $\dim(\bar{\xi(T_1)})=p\wedge q\wedge s$.
\end{itemize}
\end{dfn}

\begin{rem}\label{rdv.3.4}
In (\ref{5.13}) $F(\bar{x},\bar{z},\bar{\xi},w)$ is a  virtual force, caused by the interaction of docking objects, satisfying
$$
\Psi(\bar{x},\bar{z},\bar{\xi},w)=0,\quad if
\begin{bmatrix}
\bar{x}\\
\bar{z}
\end{bmatrix}\lvminus \bar{\xi}=0.
$$
\end{rem}


\section{Concluding Remarks}\label{s8}

The main purpose of this paper is to construct a new geometric object called the DFES (or DFM), which provides a framework (i.e., the state space) for DVDS. We briefly summarize the results.

The DFES is constructed as follows:

\begin{itemize}
\item Step 1: Define an inner product for two vectors of different dimensions. It turns the ESDD ${\cal V}=\R^{\infty}=\bigcup_{n=1}^{\infty}\R^n$ into a distance space.

\item Step 2: Two vectors $x,y\in \R^{\infty}$ are said to be equivalent, denoted by $x\lra y$, if their distance is zero. The quotient space $\Omega=\R^{\infty}/\lra$ is called the DFES. In fact, $d_{{\cal V}}(x,y)=0$, if and only if, there exist $\J_{\a}$ and $\J_{\b}$ such that
    $x\otimes \J_{\a}=y\otimes \J_{\b}$. It is clear that two vectors are equivalent if they contain the same information. In other words, vector form is a way for a set of data to express itself. It may be expressed as vectors of different dimensions, but from information point of view, they are equivalent.

\item Step 3: By posing scalar multiplication and addition $\Omega$ becomes a topological real vector space. Let $\Pr: \R^{\infty}\ra \Omega$ be the natural projection. Then $(\R^{\infty},\Pr,\Omega)$ becomes a fiber bundle.

\item Step 4. Using the fiber bundle structure of $(\R^{\infty},\Pr,\Omega)$, each $\bar{x}\in \Omega$ has a coordinate neighborhood, which is a set of coordinate charts of various dimensions. It is called a bundle of coordinate charts.  Hence $\Omega$ is called a DFES.

\item Step 5. Using the bundles of coordinate charts, a differentiable structure can be posed on $\Omega$, making it a DFM. Then the continuous functions, (co)-vector fields, (co)-distributions, and tensor fields can be built for $\Omega$. Eventually, the dimension-free Riemannian manifolds and dimension-free symplectic manifolds can be properly constructed.

\end{itemize}

Note that the gluing topology on the DFES $\Omega=\R^{\infty}/\lra$ makes it a path-wise connected topological space. Therefore, intuitively, the trajectories of dynamic systems over $\Omega$ can continuously move ``across" Euclidian spaces of different dimensions. This is the main idea for using DFM to design DVDS and DVCS. Lifting the trajectory of a dynamic system over $\Omega$ to a leaf (a Euclidian space of fixed dimension) is called a realization. As a dynamic system over $\Omega$ is lifted onto leafs of different dimensions, a dimension-varying realization is obtained. Conversely, we can also project the trajectory of a dynamic system on a  Euclidian space of fixed dimension onto $\Omega$.

The design of dimension-varying dynamic (control) systems can be described as follows:

\begin{itemize}
\item Step 1: Project a dimension-varying dynamic system, which has broken vector fields over Euclidian spaces of different dimensions, onto $\Omega$ to form a dynamic system over $\Omega$, which consists of several (finite number) of vector fields.
\item Step 2: Lifting the dynamic system on $\Omega$ to a Euclidian space of proper dimension, where all the vector fields involved by the dynamic system on $\Omega$ can be properly lifted into this Euclidian space.
\item Step 3: All the analysis and control design can be done in conventional way for this lifting system on its Euclidian space.

\item Step 4: Project the resulting manipulated system back to $\Omega$ and then lifting it into several original Euclidean spaces, where the original dimension-varying system lies on.

\end{itemize}

Finally, we would like to present a conjecture: The DFM might provide a framework (i.e.,  the state space) for string theory in physics. The idea is sketched as follows:

Consider a subspace of DFM as $\Omega_3:=\left\{\bar{x}\in \Omega\;|\;\dim(\bar{x})\leq 3\right\}$.
We choose $3$ because it is the dimension of real physical world.

Now $(\R^{\infty},Pr,\Omega_3)$ is a sub-bundle of the fiber-bundle $(\R^{\infty},\Pr,\Omega)$. If we consider all possible realization of dynamic systems over $\Omega_3$, then the minimum total subspace which allow all possible realizations is $\R^{[\cdot,6]}$.
Hence, if we desire a space which is of minimum dimension and contains all moves (or dynamic systems), then it is ${\cal B}:=(\R^{[\cdot,6]}\ra \Omega_3)$.
Since this manifold is of dimension $9$, plus a dimension for $t$, a manifold of dimension $10$ is reasonable for describing state-motion-time. This might be the string space.

Some further arguments are the following:

\begin{itemize}
\item[(i)] It is well known in classical differential geometry that an $n$-dimension manifold $M$ has an  $n$-dimensional tangent space at each point. Hence, if taking both $M$ and $T(M)$ into consideration, an $n$-dimensional manifold with its tangent bundle is a $2n$ dimensional manifold, which is a well known fact. So consider the bundle ${\cal B}$ as a $9$-dimensional manifold is reasonable.

\item[(ii)] It seems that there is no static particle in the world. That is, particles are always joined with their moves. Moves can be described by vector fields. So to describe a particle, a position plus a vector field on its tangent space may be reasonable to describe it, as the particle is small and its movement is very fast. Using string to describe a particle might essentially be a description for both the position and the trajectory of a particle.

\item[(iii)] Now taking the position and moving trajectory into consideration. We may consider the extra $6$ dimensions being used to describe open string (or open movement of a particle). In addition, we need $SU(3)$ to describe the gauge group and $SU(1)$ for rotation. Then we have
    ${\cal B}+SU(3)+SU(1)+\mbox{time}$, which is of dimension $26$. This manifold might be proper for Bosonic super-sting model.
\end{itemize}

In brief, DFM could provide a framework for systems with arbitrary dimensions. DFM with a Reimannian structure becomes a DFRM. The investigation of DFRM in this paper is very elementary. A continuous study is necessary. It is promising that DFRM  might overcome the crisis of classical Riemannian geometry \cite{zhapr}.

\end{document}